\UseRawInputEncoding
\documentclass[twocolumn]{aastex631}

\usepackage{hyperref, float, graphicx, verbatim, longtable, booktabs}
\newcommand{\Lcol}[1]{\parbox[t]{6.2cm}{#1}}  
\newcommand{\Rcol}[1]{\makebox[0.8cm][r]{#1}} 

\begin{document}

\title{Tracing Quenching in Nearby Galaxies Through Inner Surface Mass Density and Cold Gas Content}

\author[0000-0003-1211-530X]{Evangela E. Shread\textsuperscript{\textdagger}}
\affiliation{Department of Astronomy, Orange Coast College, Costa Mesa, CA 92626}
\affiliation{Division of Physics, Math and Astronomy, California Institute of Technology, Pasadena, CA 91125}

\author[0009-0008-5554-9144]{Trevor J. Weiss\textsuperscript{\textdagger}}
\affiliation{Department of Astronomy, Orange Coast College, Costa Mesa, CA 92626}
\affiliation{Department of Physics and Astronomy, California State University, Long Beach, Long Beach, CA 90840}

\author[0000-0002-7254-7174]{Jerome J. Fang}
\affiliation{Department of Astronomy, Orange Coast College, Costa Mesa, CA 92626}

\author{Cameron Law}
\affiliation{Department of Astronomy, Orange Coast College, Costa Mesa, CA 92626}
\affiliation{Department of Physics and Astronomy, University of California, Irvine, Irvine, CA 92697}

\footnotetext{\textsuperscript{\textdagger} Authors contributed equally.}

\begin{abstract}

The inner stellar mass surface density within 1 kpc, $\Sigma_1$, has emerged as a suitable proxy for bulge growth and galaxy quenching. However, the dependence of cold gas content on $\Sigma_1$ has not been thoroughly explored. In this paper, we examine the relationship between $\Sigma_1$, as well as the mass-relative parameter $\Delta\Sigma_1$, and the atomic ($f_\mathrm{HI}$) and molecular ($f_\mathrm{H2}$) cold gas fractions in massive, nearby galaxies. We utilize a sample of 341 galaxies with H{\small I} data and 201 galaxies with H$_2$ data from the xGASS and xCOLDGASS surveys, spanning $0.02\le z \le 0.05$ and a stellar mass range of $10^{10} \le M_{\ast}/M_{\odot} \le 10^{11.5}$. While we observe that a decline in both $f_\mathrm{HI}$ and $f_\mathrm{H2}$ is associated with increasing $\Sigma_1$, we find that $f_\mathrm{H2}$ shows a sharper decline above a threshold value of $\Delta\Sigma_1 = 0$. In addition, the fraction of galaxies with AGN activity (Seyferts and LINERs) increases with $\Delta\Sigma_1$, with the greatest increase occurring between $0 \lesssim \Delta\Sigma_1 \lesssim 0.2$ dex. We propose an evolutionary track in the plane of $f_\mathrm{H2}-\Delta\Sigma_1$, whereby molecular gas depletion at fixed mass coincides with a rise in AGN activity. Our results suggest that central bulge growth is more tightly coupled to the depletion of molecular gas rather than atomic gas, with AGN feedback possibly contributing to this process. Our work highlights the utility of $\Sigma_1$ and $\Delta\Sigma_1$ as tracers of quenching in massive galaxies.

\end{abstract}


\section{Introduction} \label{sec:intro}

Star formation quenching and the processes that regulate it are integral to understanding galaxy evolution. Two important markers of quenching are variations in galaxy structure and a decline in the availability of cold gas. The relationship between cold gas content and galaxy structure has been studied extensively, beginning with early works which established a connection between Hubble morphological type and global cold gas content. The H{\small I} gas fraction ${(f_{\mathrm{HI}} \equiv M_{\mathrm{HI}}/M_{\ast})}$ increases along the Hubble sequence from early- to late-type galaxies (e.g. \citealt{Roberts69, Young89, Sage93, Conselice14}). There exists less variation in the H$_2$ gas fraction $(f_{\mathrm{H2}} \equiv M_{\mathrm{H2}}/M_{\ast})$ among spiral morphological types, but elliptical types have appreciably lower fractional H$_2$ mass content and often constitute non-detections in targeted surveys (e.g. \citealt{Young91, Sage93, Roberts94, Boselli14}). 

Scaling relations between galactic structure and cold gas mass offer a means of quantifying the interplay between gas content and galaxy evolution. There have been many studies of the relationship between H{\small I} and H$_2$ content and structural parameters, such as stellar mass surface density (e.g. \citealt{Saintonge11, Catinella13, Brown15, Saintonge17, Catinella18, Baker21}); the half-light or effective radius (e.g. \citealt{Yesuf19, Lin20}); S\'ersic index (e.g. \citealt{Fisher13, Namiki21}); and non-parametric measures such as concentration index, asymmetry, and the Gini coefficient (e.g. \citealt{Catinella10, Wang11, Kauffmann12, Zhou18, Namiki21, Davis22}). 

The strongest cold gas scaling relations with galactic structure typically involve stellar mass and stellar mass surface density, defined in this work as $\mu_{\ast} \equiv 0.5M_{\ast}$/$\pi R_{50,z}^2$, where $R_{50,z}$ is the half-light radius computed from the SDSS \textit{z} band flux. Stellar mass surface density has been used to quantify the prominence of the central bulge component, whereby galaxies become bulge-dominated at $\log \mu_{\ast} > 8.7 \ M_\odot \  \mathrm{kpc}^{-2}$ (\citealt{Catinella10}). For nearby galaxies, $\mu_{\ast}$ is anti-correlated with both $f_{\mathrm{HI}}$ and $f_{\mathrm{H2}}$, such that more bulge-dominated galaxies have lower gas fractions (e.g. \citealt{Catinella10, Saintonge17}). Stellar mass is strongly anti-correlated with atomic gas fraction; $f_{\mathrm{HI}}$ scales with $\log M_\ast$ with a slope ranging from $-0.3$ to $-0.9$ dex, corresponding to a drop of $\sim70-80\%$ in the gas fraction per dex in $M_\ast$ (e.g. \citealt{Brown15, Catinella18, Hunt20}). In contrast, molecular gas fraction shows a weak dependence with $M_\ast$, with previous studies finding a near-flat relation, e.g. a median of $\log f_{\mathrm{H_2}}\simeq -0.99 \pm 0.22$ and a slope consistent with zero over a broad mass range (\citealt{Saintonge11, Saintonge17}).

The focus of this paper is to examine the relationship between cold gas content and the stellar mass surface density within 1 kpc of the galactic center ($\Sigma_1$) in nearby, massive galaxies. While similar to $\mu_{\ast}$, $\Sigma_1$ measures the stellar mass surface density within a smaller, constant radius. $\Sigma_1$ and similar parameters (such as the inner surface mass density within 1.5 or 2 kpc) have become popular metrics in recent years to identify and examine galaxies that have been quenched. Early works presented $\Sigma_1$ as a proxy for central bulge growth and for central black hole mass ($M_\mathrm{BH}$), and have proposed that once galaxies reach a mass-dependent $\Sigma_1$ threshold value, their star formation is shut down (\citealt{Cheung12, Fang13, Barro17}). For this reason, $\Sigma_1$ is often used as a tracer of galactic evolution from the star-forming to the quiescent phase. As $\Sigma_1$ grows at fixed mass, galaxies transition through the green valley from the blue sequence to the red (\citealt{Fang13}), and at fixed redshift, galaxies evolve along the star formation main sequence (MS) via inside-out growth, traced by increasing $\Sigma_1$ (\citealt{Barro17}). 

$\Sigma_1$ is tightly correlated with $M_\ast$ (e.g. \citealt{Fang13}). The mass-trend-removed parameter $\Delta\Sigma_1$ is defined relative to the \textit{structural valley} within the $\Sigma_1-M_\ast$ plane. This is a visual feature in the plot which, at fixed mass, appears as a trough between two higher density populations of low and high $\Sigma_1$ (\citealt{Luo20}). While not exactly the same as the star-formation green valley, \cite{Luo20} showed that $\Delta\Sigma_1$ as defined in this manner is related to the star-forming/quiescent division in massive galaxies. \cite{Luo20} calculated a parabolic fit to mark the structural valley in the $\Sigma_1-M_\ast$ plane, and $\Delta\Sigma_1$ is defined as the offset in $\log\Sigma_1$ from this line at fixed mass. With the mass trend removed, $\Delta\Sigma_1$ provides a more robust probe of the relationship between inner stellar mass surface density growth and global cold gas fraction.

When $\Sigma_1$ or $\Delta\Sigma_1$ is plotted against SFR, or alternatively against other parameters that trace SFR--such as color, specific SFR (sSFR), or distance from the MS--a distinctive `elbow' or `L-shaped' pattern emerges in the distribution of galaxies in bins of $M_\ast$ (\citealt{Fang13, Barro17, Lee2018, Luo20, Guo2021}). It has been proposed that there is more than one possible track along which galaxies evolve as they transition from star-forming to quiescent. For example, in the plane of $\Sigma_1$ vs. sSFR, evolution may be characterized by either gas-rich or gas-poor quenching processes, which correspond to quenching at earlier and later times, respectively (\citealt{Barro13, Barro17}). For satellites, differences in evolution within this plane may involve quenching via bulge growth driven by central star formation, or through global mass-loss processes like tidal stripping (\citealt{Woo16}). In any case, the consensus of the $\Sigma_1$ literature is that galaxies must build up their central bulge as a prerequisite for quenching.

The proposed mechanisms of massive galaxy quenching are too complex to discuss in depth here, and vary according to a variety of galaxy properties. Theoretical and simulation analyses contend that AGN feedback is a prominent -- even necessary -- mechanism for massive galaxy quenching (e.g. \citealt{Croton_2006, Dubois2013, Su2019, Piotrowska2021}). Because of the close relationship between $\Sigma_1$ and the growth of the central black hole, we focus on the role of AGN feedback in gas removal and quenching and its connection to $\Delta\Sigma_1$.

It has been established that $f_\mathrm{H2}$ is tightly correlated with sSFR (\protect\citealt{Saintonge17}). In the plane of sSFR and $\Delta\Sigma_1$, galaxies follow a bimodal distribution which approximately separates star-forming from quiescent galaxies (\protect\citealt{Luo20}). These two relationships motivate us to investigate whether central bulge growth is also related to global cold gas content, particularly H$_2$. Moreover, the scaling of cold gas content with $\Sigma_1$ has not been thoroughly explored in previous research. A few studies have analyzed the relationship between $\Sigma_1$ and cold gas content within the framework of cosmological simulations. \cite{Ma2022} and \cite{Khan_2025} each concluded that $\Sigma_1$ is connected to a decline in cold gas in galaxy centers within the IllustrisTNG and NIHAO simulations, respectively, and offered AGN feedback as a plausible explanation. In this work, we will use observational data to compare the relationship between $\Sigma_1$ and global cold gas content to established gas scaling relations with $M_\ast$ and $\mu_\ast$. With the availability of molecular and atomic cold gas catalogs for representative samples, we are also positioned to explore the usefulness of $\Sigma_1$ and $\Delta\Sigma_1$ as indicators of galaxy quenching through an analysis of the cold gas reservoirs of galaxies with a range of $\Delta\Sigma_1$ values.

In Section \ref{sec:sample}, we discuss our sample selection from archival data. Section \ref{sec:results} presents our analysis of the relationship between $\Sigma_1$, AGN activity and gas content in our sample. In Section $\ref{sec:discussion}$, we discuss our findings that $\Sigma_1$ is correlated with cold gas fraction above a certain threshold. We then connect the observed distribution of objects in the plane of $\Delta\Sigma_1$ vs. $f_\mathrm{H2}$ to existing observational and theoretical explanations for how galaxies evolve from star-forming to quiescent. Section $\ref{sec:conclusion}$ summarizes our results. Throughout this work, we assume a concordance $\Lambda$CDM cosmology: $H_0 = 70 \ \mathrm{km \ s^{-1} \ Mpc ^{-1}}$, $\Omega_m = 0.3$, $\Omega_\Lambda = 0.7$.

\section{Sample Selection} \label{sec:sample}
The Extended GALEX Arecibo SDSS Survey (xGASS; \citealt{Catinella18}) and Extended CO Legacy Database for GASS (xCOLDGASS; \citealt{Saintonge17}) provide publicly available H{\small I} and H$_2$ catalogs for mass-selected samples of nearby galaxies. Spectral measurements were obtained from the MPA/JHU DR7 catalog\footnote{\href{https://wwwmpa.mpa-garching.mpg.de/SDSS/DR7/}{https://wwwmpa.mpa-garching.mpg.de/SDSS/DR7/}} for the purposes of BPT classification (\citealt{Brinchmann04}). The samples were then combined with $\Sigma_1$ data (\citealt{Luo20}).

\begin{figure*}
    \centering
    \includegraphics[width=\linewidth]{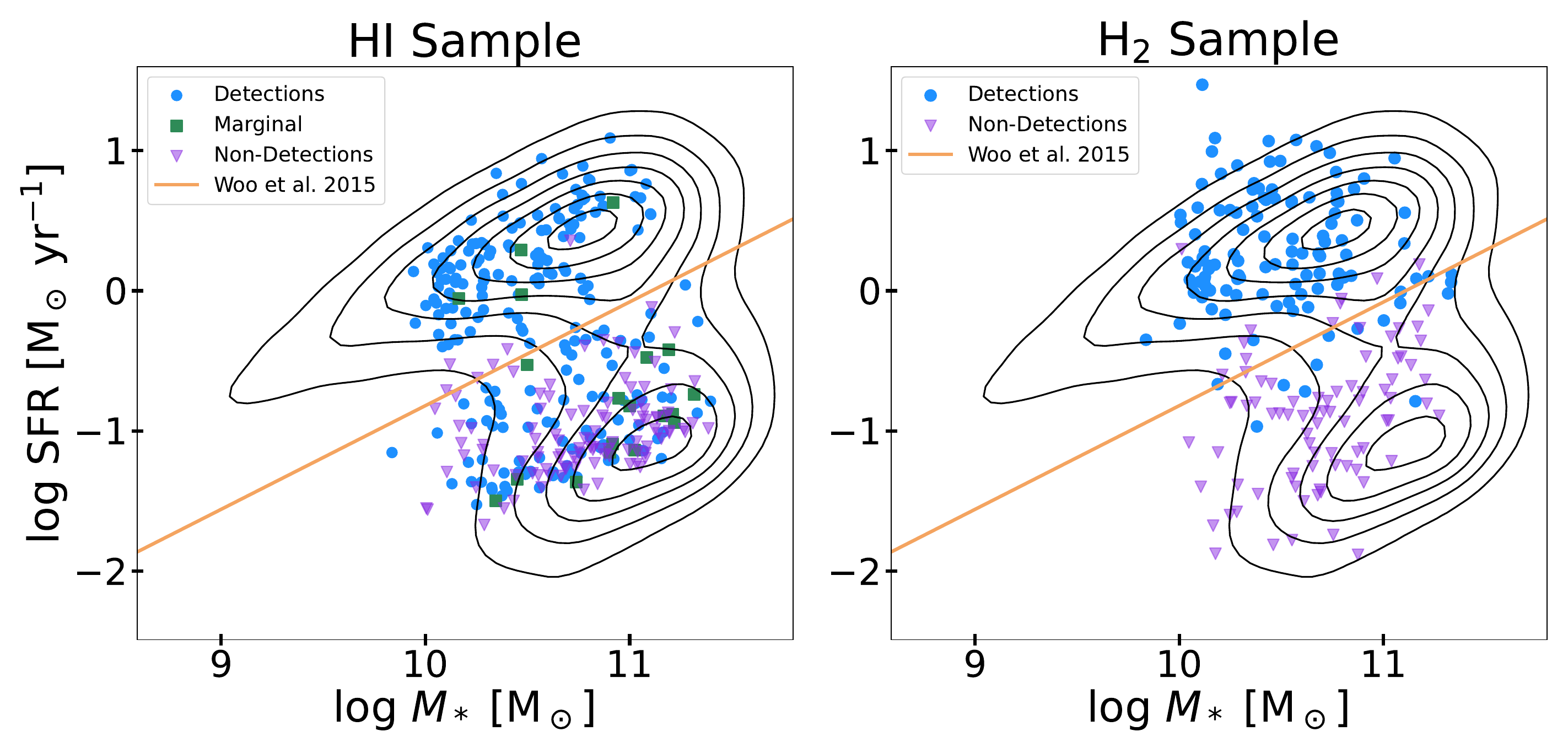}
    \caption{(\textit{Left}) The contours represent the SDSS main sample on the SFR-$M_{\ast}$ plane, and we include the line from \citealt{Woo15} to indicate objects that are either star-forming (above the line) or non-star-forming (below the line). The points comprise our final H{\small I} sample from the original xGASS sample. Code 2 xGASS detections are square markers to distinguish them as ``less reliable" or ``marginal" detections. (\textit{Right}) The contours of the full SDSS sample are now overlaid with the points comprising our H$_2$ sample, which was derived from the full xCOLDGASS sample. In both plots, non-detections are indicated by triangle markers. For further explanation, see Sections \ref{subsec: HI Sample} and \ref{subsec: H2 Sample}.}
    \label{fig:contour}
\end{figure*}

In Figure \ref{fig:contour}, we show the distribution of our H{\small I} and H$_2$ samples in the SFR-$M_\ast$ plane, compared to the SDSS DR7 Catalog (\citealt{Abazajian09}). To identify galaxies as either star-forming or quiescent, we adopted the criterion set by \cite{Woo15}, who divided SDSS galaxies in the SFR-$M_\ast$ plane according to the line $\log \text{SFR} = 0.74 \log M_\ast - 8.22$. Galaxies above this line are classified as star-forming, and those below it are considered quiescent. This line also fits our data well, even though the SFR values we use originate from a different source (the xGASS and xCOLDGASS catalogs).

\begin{table}[t!]
    \begin{tabular}{@{}l r@{}}
    \multicolumn{2}{c}{\textbf{xGASS H{\small I} Sample}} \\
    \hline
    \Lcol{Original xGASS Sample Size} & \Rcol{$1179$}\\
    \Lcol{Cross-Matched with $\Sigma_1$ Catalog (\protect\citealt{Luo20})} & \Rcol{$448$}\\
    \Lcol{Emission-line fluxes with S/N $> 3$} & \Rcol{$370$}\\
    \Lcol{Excluding code 3 or higher (i.e., confused) xGASS detections} & \Rcol{$341$}\\
    \hline 
    \Lcol{\textbf{Final Sample Size}} & \Rcol{\textbf{341}} \\
    \Lcol{\hspace{0.2cm} Reliable Detections} & \Rcol{$213$}\\
    \Lcol{\hspace{0.2cm} Marginal Detections} & \Rcol{$19$}\\
    \Lcol{\hspace{0.2cm} Non-detections} & \Rcol{$109$}\\
\end{tabular}
\caption{The original xGASS survey sample with detailed cuts made to the data set. The relative partitioning of our sample into reliable detections, marginal detections, and non-detections, or upper limits, is quantified.}
\label{table:gass}
\end{table}

\begin{table}[t]
    \begin{tabular}{@{}l r@{}}
    \multicolumn{2}{c}{\textbf{xCOLDGASS H$_2$ Sample}} \\
    \hline
    \Lcol{Original xCOLDGASS Sample Size} & \Rcol{$532$}\\
    \Lcol{Cross-Matched with $\Sigma_1$ Catalog (\protect\citealt{Luo20})} & \Rcol{$226$}\\
    \Lcol{Emission-line fluxes with S/N $> 3$} & \Rcol{$201$}\\
    \hline
    \Lcol{\textbf{Final Sample Size}} & \Rcol{\textbf{201}}\\
    \Lcol{\hspace{0.2cm} Detections} & \Rcol{$118$}\\
    \Lcol{\hspace{0.2cm} Non-detections} & \Rcol{$83$}\\
\end{tabular}
\caption{The original xCOLDGASS survey sample with detailed cuts made to the data set. The distribution between detections and non-detections is quantified.}
\label{table:coldgass}
\end{table}

\subsection{\texorpdfstring{$\Sigma_1$}{Sigma1} Data}
\cite{Luo20} measured $\Sigma_1$ values for a volume- and mass-limited sample of $>40,000$ SDSS galaxies. In particular, they focused on central galaxies with good viewing geometries, imposing the following restrictions: stellar mass $M_\ast >10^{9.5} M_\odot$; axis ratio $b/a > 0.5$; S\'ersic index $0.5 < n < 6$; and redshift $0.02 < z < 0.07$. See \cite{Luo20} for details of their survey methods. 

With the above restrictions, cross-matching our gas data with the $\Sigma_1$ data reduced our H{\small I} sample from the 1179 objects comprising the full xGASS catalog to 448, and reduced our H$_2$ sample from the original 532 objects of the xCOLDGASS sample to 226.

Since we are using the $\Sigma_1$ values calculated by \cite{Luo20}, we use the same stellar mass data for consistency, which are obtained from the MPA/JHU DR7 value-added catalog. We also adopt their computed $\Delta\Sigma_1$ values in our analysis. 

\subsection{H{\small I} sample}
\label{subsec: HI Sample}
xGASS is a targeted survey of galaxies with $10^{9} M_\odot < M_\ast < 10^{11.5} M_\odot$ and redshifts $0.01 < z < 0.05$. Integration times were set by the requirement that certain upper limits be reached unless a measurement could be obtained with a sufficient S/N ratio, allowing for greater sensitivity in the low $f_\mathrm{HI}$ regime. The distribution of galaxies is approximately uniform in stellar mass bins between $\log M_\ast/M_\odot = 10-11$, and decreases steadily for $\log M_\ast/M_\odot > 11$. The representative sample contains 1179 galaxies, a fraction of which are galaxies with archival H{\small I} data obtained from ALFALFA or the Cornell H{\small I} Archive, which were not remeasured for survey efficiency. For further information on how H{\small I} data were collected and chosen for the xGASS sample, see \cite{Catinella18}, and for general GASS survey methods, see \cite{Catinella10} and \cite{Catinella13}.  

For the purposes of this work, we consider both H{\small I} detections and non-detections. Non-detections are listed with 5$\sigma$ upper limits in the xGASS tables. If measured by GASS, the highest quality, or ``code 1'' detections, were detected with S/N $\gtrsim 6.5$ and are considered reliable (\citealt{Catinella18}). 
Code 3 or higher H{\small I} detections are flagged by \cite{Catinella18} as being ``marginal and confused'' or simply ``confused". For these targets, H{\small I} emission is believed to originate from a different source within the beam. There are a total of 74 such measurements in the full xGASS sample. 

Code 2 detections are marginal in that they have lower S/N (between 5 and 6.5) compared to code 1 detections. They are \textit{not} confused, and as such they can be considered secure detections, albeit with more uncertain H{\small I} masses. In our figures, code 2 detections are labeled separately from the sample of code 1 xGASS detections and high-confidence archival measurements. There are 19 objects in our sample that fall into this category. The motivation to include these points graphically is to provide more coverage of the parameter space. However, in calculating sample statistics such as averages, these points are excluded.

\begin{figure*}
    \centering
    \includegraphics[width=\linewidth]{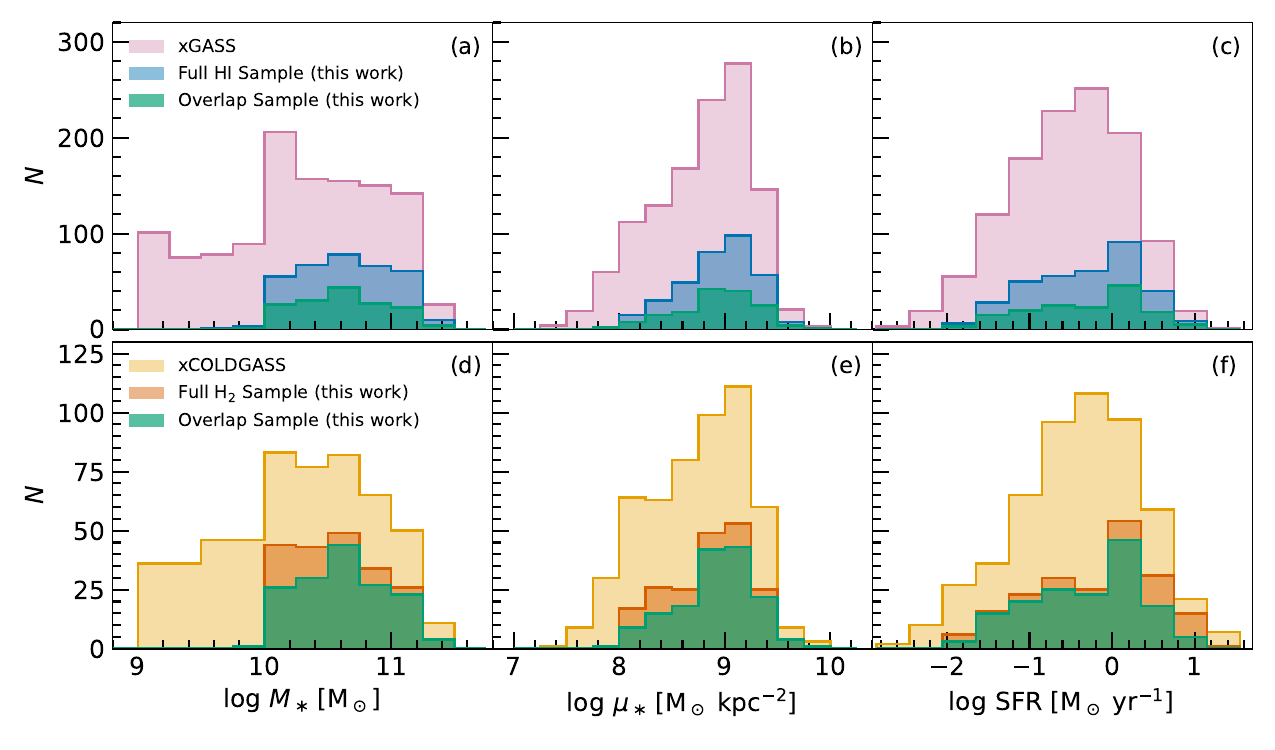}
    \caption{The distributions of (a) $M_\ast$, (b) $\mu_\ast$ and (c) SFR are compared between the xGASS sample, our H{\small I} sample, and the overlap sample, which consists of the 155 galaxies common to our H{\small I} and H$_2$ samples. (d-f) The distributions of these three key properties are compared between the xCOLDGASS sample, our H$_2$ sample, and the overlap sample.}
    \label{fig:hist_gass_coldgass}
\end{figure*}

Table \ref{table:gass} offers an analysis of all cuts made to the original H{\small I} xGASS sample, for a final sample size of 341. We compare the distributions of $M_\ast$, $\mu_\ast$ and SFR within our sample to the xGASS sample in Figure \ref{fig:hist_gass_coldgass}. Figure \ref{fig:contour} shows the distribution of these objects in the SFR-$M_\ast$ plane. As expected, galaxies along the MS are largely detected in H{\small I}, while galaxies lacking H{\small I} detections cluster in the passive sequence.

\subsection{H\texorpdfstring{$_2$}{2} sample}
\label{subsec: H2 Sample}
We extract our H$_2$ data from the xCOLDGASS sample, which spans a mass of $10^{9} < M_\ast/M_\odot < 10^{11.5}$ and redshift range of $0.01 < z < 0.05$ (\citealt{Saintonge17}). The methods of xCOLDGASS are complementary to those of xGASS, and there are 477 galaxies common to both samples. 

In xCOLDGASS, H$_2$ masses are extrapolated from CO(1-0) emission line flux measurements, which serve as a tracer for molecular gas. For details about the sampling and detection methods of xCOLDGASS, see \cite{Saintonge17}. Non-detections are reported with $3\sigma$ upper limits in the xCOLDGASS tables, and all detections have S/N $>5$ by survey design. Within our sample, there are 118 detections and 83 non-detections. 

Table \ref{table:coldgass} offers an analysis of the cuts we made to our sample, for a final sample size of 201. We compare the distributions of $M_\ast$, $\mu_\ast$ and SFR within our H$_2$ sample to the xCOLDGASS sample in Figure \ref{fig:hist_gass_coldgass}. The distribution of these objects in the SFR-$M_\ast$ plane is shown on the right in Figure \ref{fig:contour}. As with our H{\small I} sample, passive galaxies predominantly lack detections, while nearly all star-forming galaxies are detected in H$_2$.

\subsection{Our Samples in the BPT Diagram}\label{sec:BPT_diagram}
A central focus of this paper is on quantifying the relationship between AGN activity, $\Delta\Sigma_1$ and cold gas fractions. Emission-line fluxes and their errors for [O{\small III}], [N{\small II}], H$\alpha$ and H$\beta$ were extracted from the MPA/JHU DR7 catalog of spectral measurements for the purposes of BPT classification (\citealt{Baldwin81}). We impose the condition that S/N $> 3$ for each line. Upon enforcing this restriction, our samples were reduced to 370 objects with H{\small I} data and 201 objects with H$_2$ data.

\begin{figure*}[ht!]
    \centering
    \includegraphics[width=\textwidth]{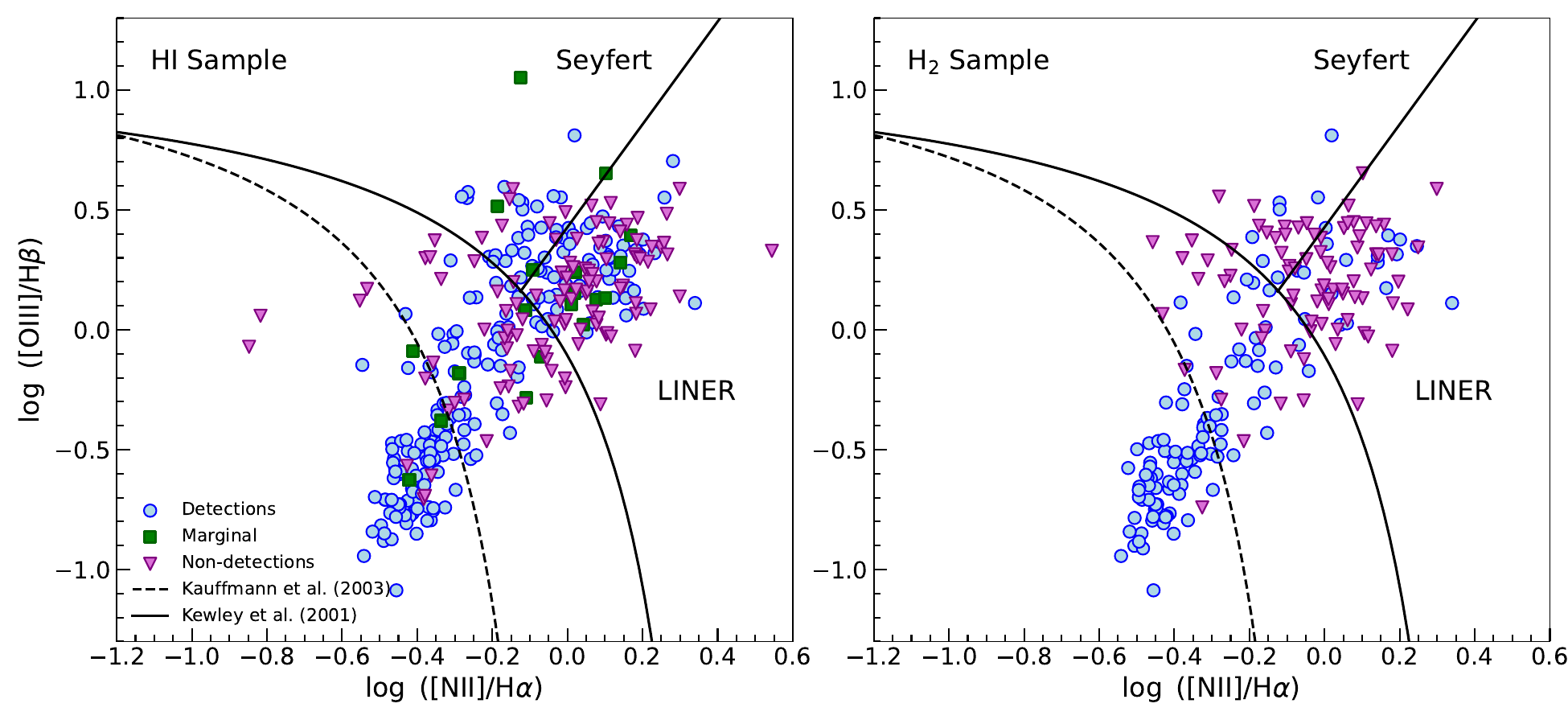}
    \caption{(\textit{Left}) The BPT diagram for our H{\small I} sample, and (\textit{right}) for our H$_2$ sample. We show the theoretical relation of \cite{Kewley01} and empirical relation of \cite{Kauffmann03}, which separate galaxies based on their dominant source of gas ionization.}
    \label{fig:BPTDiagram}
\end{figure*}

Figure \ref{fig:BPTDiagram} shows the distribution of our sample in the BPT diagram. There are fewer star-forming galaxies, demarcated by their location below the \cite{Kauffmann03} line, than typically observed for larger, more representative samples. There is a documented mass trend within the BPT diagram whereby the star-forming branch is dominated by lower-mass galaxies (e.g. \citealt{Kumari21}); since our sample is limited to higher-mass objects ($\log M_\ast/M_\odot \gtrsim 10$), the distribution of star-forming galaxies is lacking, particularly for $\log (\text{[NII]/H}\alpha) < -0.6$. 

Galaxies located above the \cite{Kewley01} line are classified as AGN candidates. Our H{\small I} sample contains 163 such candidates ($48 \%$ of the sample) and our H$_2$ sample contains 85 ($42 \%$). We note that AGN are somewhat overrepresented in our sample compared to the \cite{Luo20} sample, in which $\sim 35 \%$ of galaxies meeting their selection criteria are AGN for $M_\ast > 10^{10} M_\odot$ and $0.02 \leq z \leq 0.05$. This is discussed further in Section \ref{subsec:results_agn}.

\begin{figure*}[h]
    \centering
    \includegraphics[width=\linewidth]{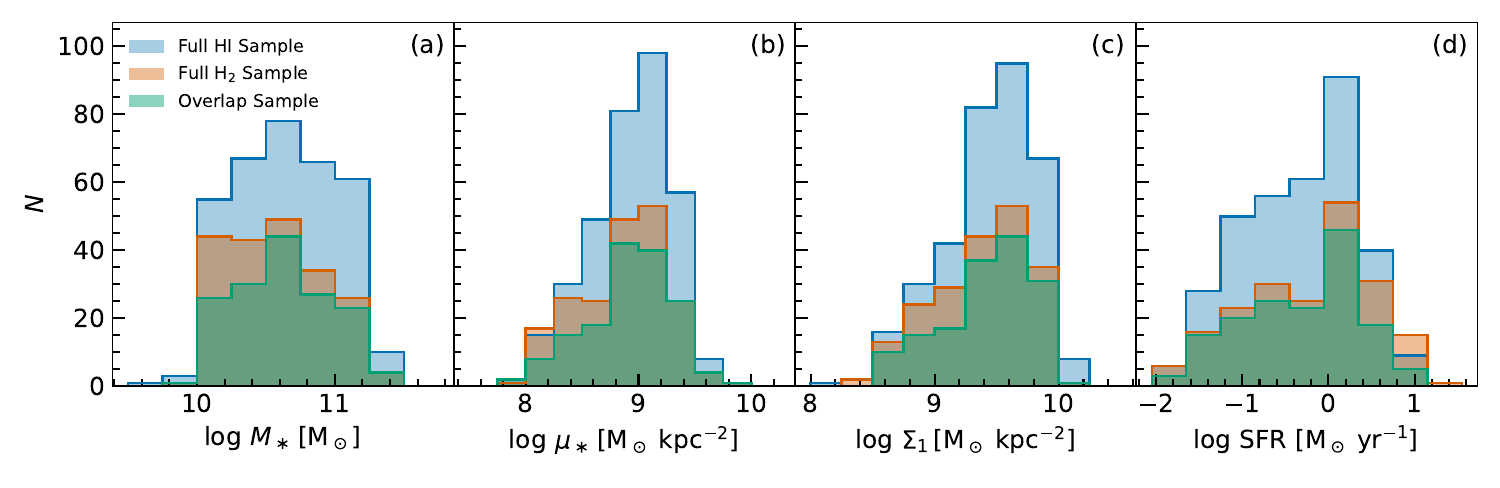}
    \caption{We show the distributions of (a) $M_\ast$, (b) $\mu_\ast$, (c) $\Sigma_1$, and (d) SFR for our H{\small I} and H$_2$ samples. The distributions are not identical for these four parameters, so we utilize the overlap sample consisting of the galaxies common to both of our samples to examine the robustness of our results. This is particularly relevant for our investigations of $\Sigma_1$ and $\Delta\Sigma_1$ vs. cold gas fraction.}
    \label{fig:hist_HI_H2_overlap}
\end{figure*}

\subsection{Overlap Sample}

Because our H{\small I} and H$_2$ samples are drawn from two surveys with different selection criteria, it is important to determine the extent to which our two samples exhibit similar distributions in key parameters. This is done to assess the robustness of our main results against potential sample biases. There are 155 galaxies that are common to both samples (the ``overlap sample''). Of these, 103 have H{\small I} detections and 86 have H$_2$ detections.

Figure \ref{fig:hist_gass_coldgass} presents the distributions in $M_\ast$, $\mu_\ast$, and SFR for our H{\small I} and H$_2$ samples and compares them with the distributions of the parent xGASS and xCOLDGASS samples. Our samples are consistent with the underlying xGASS and xCOLDGASS samples, indicating that our selection criteria have not introduced any biases in these parameters. 

Also shown in Figure \ref{fig:hist_gass_coldgass} are the distributions for the overlap sample. In general, the overlap sample has similar distributions to our H{\small I} and H$_2$ samples. One notable discrepancy is in stellar mass. This is made clearer in Figure \ref{fig:hist_HI_H2_overlap}, which compares the distributions for the H{\small I}, H$_2$, and overlap samples directly. More than 75\% of galaxies in our full H$_2$ sample are also in the H{\small I} sample, but we find that our full H$_2$ sample is slightly biased toward lower masses compared to both the H{\small I} and overlap samples. This can be attributed to the differing distributions in $M_\ast$ between the xGASS and xCOLDGASS samples (Figure \ref{fig:hist_gass_coldgass}). We also note that galaxies with masses between $10 < \log(M_\ast / M_\odot) < 10.25$ appear to be underrepresented in our H{\small I} sample compared to the xGASS sample (Figure \ref{fig:hist_gass_coldgass}). 

To further quantify any differences between our H{\small I} and H$_2$ samples, we use the two-sample Kolmogorov-Smirnov (K-S) statistic to estimate similarity in the distributions of $M_\ast$, $\mu_\ast$, $\Sigma_1$, and SFR between our two samples. Because the significant overlap between our samples violates the assumption of independence in the na{\"i}ve application of the K-S test, we perform a permutation test on the K-S statistic that preserves the shared data and randomly reassigns the non-shared data \citep{Praestgaard1995}. We calculate the K-S statistic ($D$) and its empirical $p$ value after $10,000$ permutations. We choose a significance level of $\alpha = 0.05$ for testing the null hypothesis that the H{\small I} and H$_2$ distributions are identical for the property in consideration.

We find no statistically significant differences between the H{\small I} and H$_2$ distributions of SFR ($p = 0.10$) and $\mu_\ast$ ($p = 0.12$). We find statistically significant, but \textit{small} differences between the H{\small I} and H$_2$ distributions for $\Sigma_1$ ($D = 0.10$, $p = 0.001$) and $M_\ast$ ($D=0.10$, $p = 0.003$). These differences notwithstanding, we show below that they do not affect our results.

\begin{figure*}[ht!]
    \includegraphics[width=\textwidth]{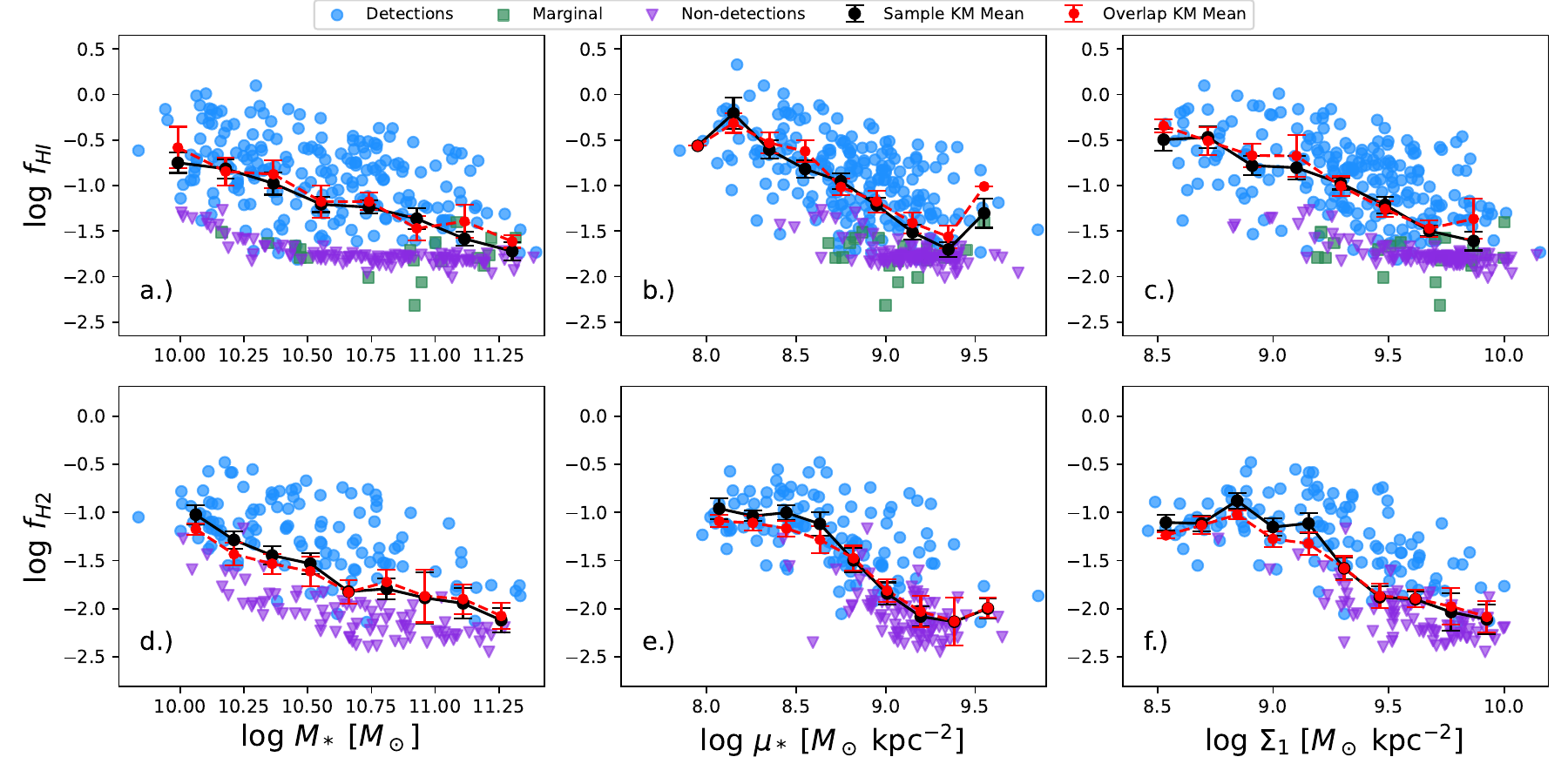}
    \caption{Gas fractions $f_{\mathrm{HI}}$ and $f_{\mathrm{H2}}$ plotted against $\Sigma_1$, $M_\ast$, and $\mu_\ast$. \emph{(Top Row)} xGASS H{\small I} detections are blue circular markers, marginal detections are green square markers, and non-detections, or upper limits, are purple triangular markers. \emph{(Bottom Row)} Blue circular points represent xCOLDGASS H$_2$ detections and purple triangular markers indicate upper limits. Black markers indicate the Kaplan-Meier means of the individual samples (non-detections and detections, excluding marginal ones), while the red markers indicate the Kaplan-Meier means of the overlap sample.}
    \label{fig:gas_fraction}
\end{figure*}

\section{Results} \label{sec:results}
\subsection{Statistical Methods}
Since our samples include objects with low SFR (and presumably low gas content), it is important that we properly treat objects in our samples that have non-detections in H{\small I} and/or H$_2$. It is common to use statistical methods such as survival analysis or spectral stacking to robustly treat upper limits for the purposes of including them in data analysis. For a recent overview of such methods in astronomy, see \cite{Saintonge22}. Within bins of $\Sigma_1$ or $\Delta\Sigma_1$, the detection rate was as low as 20\%. The median of the distribution is not meaningful in such cases. The Kaplan-Meier (KM) estimator, a non-parametric statistic used in survival analysis, can provide feasible sample statistics down to a detection rate of approximately 20\%, albeit with greater uncertainty. In order to include upper limits in our analysis, we elect to use the KM estimator to compute means in the gas fraction within bins of the independent variable in our analyses, adhering to the methods of \cite{Feigelson85}. We simulated uncertainty in the gas fraction by bootstrapping; the data was resampled 1000 times with replacement, and the survival function and resulting binned average gas fractions were recomputed for each trial. We use the 5th and 95th percentiles in the distribution of gas fraction means for the 1000 trials to assign a measure of confidence to the ``true" gas fraction means estimated via the KM estimator.

Quantifying the slope or scatter of a linear trend is also challenging in the presence of upper limits, when traditional methods are no longer valid. Consequently, we apply the Bayesian hierarchical model for linear regression implemented in the IDL procedure \verb|LINMIX_ERR| (\citealt{LINMIX2007}). This method is capable of fitting a straight line to data which includes non-detections in the $y$ variable, making it a suitable choice for our study. In addition to linear regression parameters, \verb|LINMIX_ERR| returns the Gaussian intrinsic scatter of the dependent variable about the regression line. This method has been used widely in astronomy (e.g. \citealt{Bentz_2013, McConnell_Ma_2013}), and it has recently been applied to xCOLDGASS data (\citealt{Hagedorn2024}). For details about the method itself, see \cite{LINMIX2007}.

In the remainder of this section, the dependence of cold gas content on $\Sigma_1$ is compared with the dependencies on $\mu_{\ast}$ and $M_{\ast}$, as these parameters have well-established relationships with cold gas content. We then examine the connection between $\Delta\Sigma_1$, gas fraction, and BPT classification. In all plots, a minimum of 10 points is required in each bin.

\begin{figure*}[ht!]
    \centering
    \includegraphics[width=\linewidth]{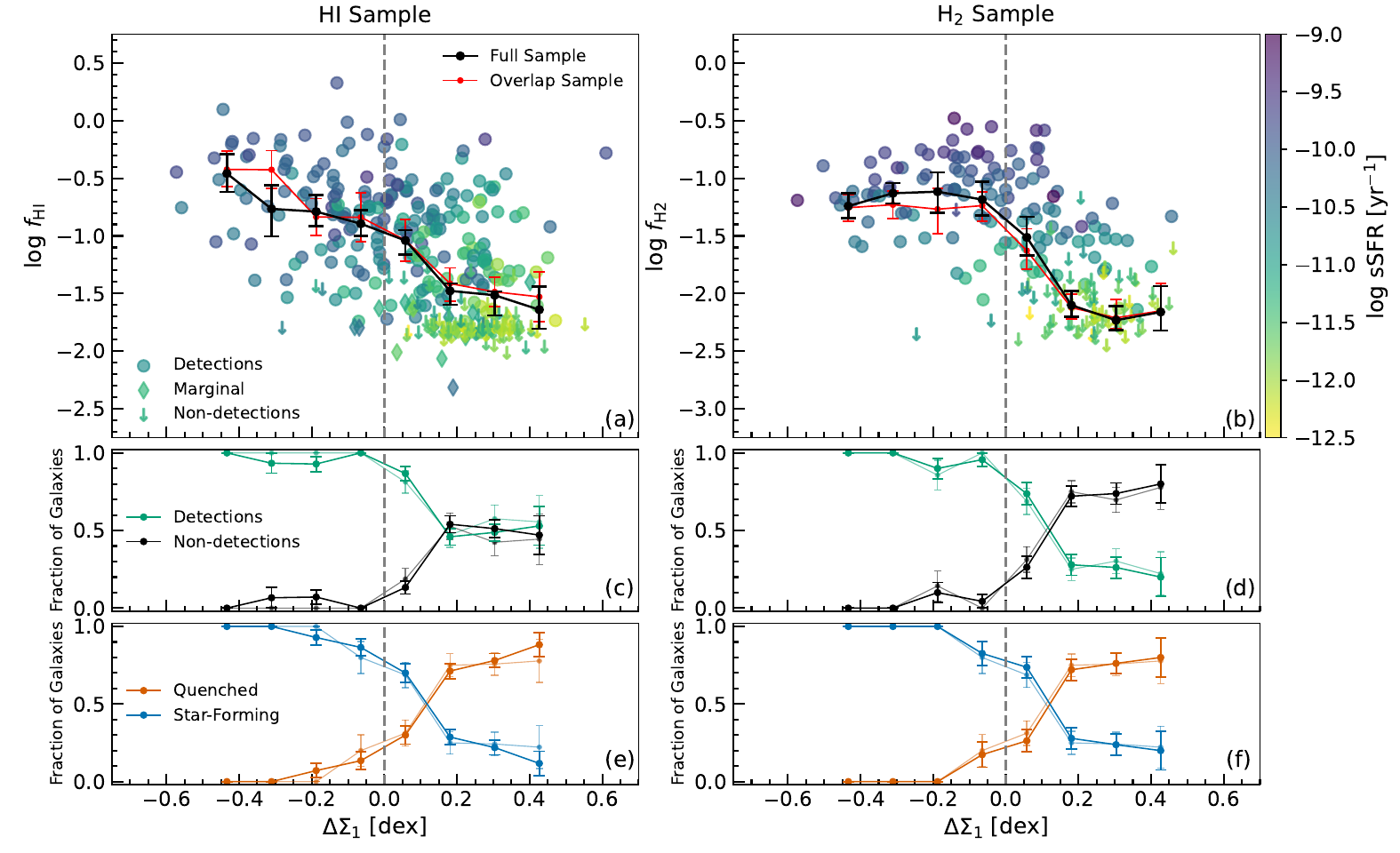}
    \caption{The dependence of (a) $f_\mathrm{HI}$ and (b) $f_\mathrm{H2}$ on the relative parameter $\Delta\Sigma_1$, coded by specific SFR. The black points represent the mean  $f_\mathrm{HI}$ and $f_\mathrm{H2}$ values computed using survival analysis techniques within bins of $\Delta\Sigma_1 \approx 0.12$ dex. The red points represent the mean values within the overlap sample. Error bars are the 5th and 95th percentiles in the bootstrap distribution as described at the beginning of Section \ref{sec:results}. (c-d) For both samples, we plot the fraction of galaxies with gas detections and the fraction without as a function of $\Delta\Sigma_1$, relative to the total number of galaxies within a given bin of $\Delta\Sigma_1$. For (c) the H{\small I} sample, we count marginal detections as detected. (e-f) The fraction of quenched and star-forming galaxies as a function of $\Delta\Sigma_1$. Galaxies are defined as star-forming or quenched according to the SFR-$M_\ast$ line defined by \citealt{Woo15}. In plots (c-f), error estimates are obtained via bootstrapping. The thinner, transparent lines indicate the relevant fractions in the overlap sample. The trends shown in (c-f) for the full samples are identical (within the uncertainties) to those found using the overlap sample.}
    \label{fig:delSigma1_sSFR}
\end{figure*}

\begin{figure*}
    \includegraphics[width=\linewidth]{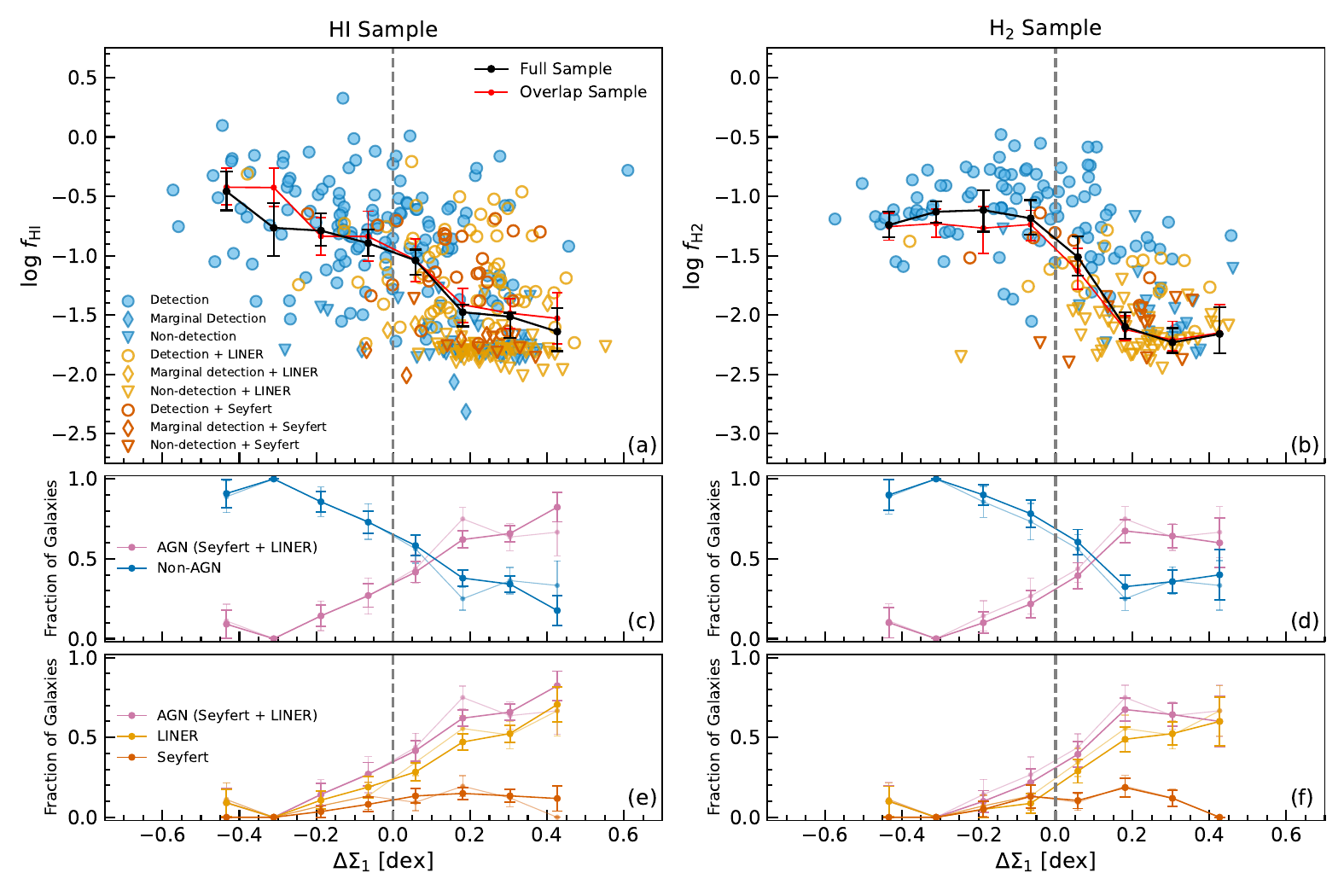}
    \caption{The dependence of (a) $f_\mathrm{HI}$ and (b) $f_\mathrm{H2}$ on the relative parameter $\Delta\Sigma_1$, coded by Seyfert or LINER classification. The black points represent the mean $f_\mathrm{HI}$ and $f_\mathrm{H2}$ values computed using survival analysis techniques. Open red markers are Seyfert candidates and open orange markers are LINER candidates selected based on their position in the BPT diagram. Blue points represent those galaxies not classified as AGN. (c-d) The fraction of galaxies within bins of $\log\Delta\Sigma_1$ that are classified as AGN candidates and the fraction of galaxies that are not are plotted for our (c) H{\small I} sample and our (d) H$_2$ sample. (e-f) The fraction of galaxies that are classified as Seyfert or LINER as a function of $\Delta\Sigma_1$. For reference, the total AGN fraction per bin is also shown. Error estimates in (c-f) are obtained via bootstrapping. The thinner, transparent lines in (c-f) indicate the relevant fractions in the overlap sample. The trends shown for the full samples are identical (within the uncertainties) to those found using the overlap sample.}
    \label{fig:delSigma1_AGN}
\end{figure*}

\subsection{\texorpdfstring{$\Sigma_1$}{Sigma1} and Cold Gas Fraction}\label{subsec:results_sigma1}
In the top row of Figure \ref{fig:gas_fraction}, we plot $f_{\mathrm{HI}}$ vs. stellar mass and structural parameters $\mu_\ast$ and $\Sigma_1$. We confirm previously established findings that $f_{\mathrm{HI}}$ is negatively correlated with $M_\ast$ and $\mu_\ast$ for nearby galaxies (e.g. \citealt{Catinella10, Huang12, Kauffmann12, Brown15, Catinella18, Hunt20}). We find that there is a similarly strong dependence of $f_{\mathrm{HI}}$ on $\Sigma_1$. 

In agreement with \cite{Saintonge17}, we observe negative correlations between $f_{\mathrm{H2}}$ and both $M_\ast$ and $\mu_\ast$ in the bottom row of Figure \ref{fig:gas_fraction}. We similarly note considerable scatter in the dependence of $f_{\mathrm{H2}}$ on $M_\ast$, with an intrinsic scatter of approximately $0.43$ dex. 

The trend between $\Sigma_1$ and $f_\mathrm{H2}$ is similar to the observed trend between $\mu_\ast$ and $f_\mathrm{H2}$, which intuitively makes sense because of the close relation between these measurements. A mild `elbow' shape emerges in both planes, suggesting nonlinear behavior. The H$_2$ fraction, as measured by the KM means, exhibits little dependence on $\Sigma_1$ when $\Sigma_1 < 10^9 \ M_\odot \ \mathrm{kpc}^{-2}$, but for $\Sigma_1 > 10^9 \ M_\odot \ \mathrm{kpc}^{-2}$, $\log f_\mathrm{H2}$ decreases almost linearly with $\log \Sigma_1$. In this regime, we find that the relationship between $\log f_\mathrm{H2}$ vs. $\log\Sigma_1$ can be described by a line with a slope of $-1.32 \pm 0.08$ and an intrinsic scatter of $0.38 \pm 0.01 \ \mathrm{dex}$. Similarly, the mean value of $\log f_\mathrm{H2}$ decreases by about $1$ dex in the range of $\mu_\ast \sim 10^{8.5} - 10^{9.5} \ M_\ast \ \mathrm{kpc}^{-2}$. This is consistent with the expected decline in molecular gas in bulge-dominated galaxies. 

We find that $\log f_\mathrm{HI}$ also decreases linearly with $\log \Sigma_1$ for $\Sigma_1 > 10^9\ M_\odot \ \mathrm{kpc}^{-2}$, with a slope of $-1.11 \pm 0.07$ and an intrinsic scatter of $0.42 \pm 0.01 \ \mathrm{dex}$. Interestingly, $\log f_\mathrm{HI}$ vs. $\log\Sigma_1$ does not exhibit nonlinear behavior to the extent we observe for $\log f_\mathrm{H2}$ vs. $\log\Sigma_1$. We elaborate on this observation in Section \ref{discussionsigma1}.

\subsection{\texorpdfstring{$\Delta\Sigma_1$}{DeltaSigma1} and Cold Gas Fraction} \label{results_delSigma1}

In Figure \ref{fig:delSigma1_sSFR}, we show the distribution of our samples in the plane of gas fraction vs. $\Delta\Sigma_1$ color-coded by sSFR. The parameter $\Delta\Sigma_1$ is defined as the offset in $\log \Sigma_1$ relative to the ``structural valley", a feature in the plot of $\log \Sigma_1$ vs. $\log M_\ast$ which generally separates bulge-dominated from disk-dominated galaxies at fixed mass (\citealt{Luo20}). The bin widths are approximately $0.12$ dex in $\Delta\Sigma_1$ in both samples. 

We observe an `elbow' pattern in the plane of $f_\mathrm{H2}$ vs. $\Delta\Sigma_1$, a shape that often emerges in the planes of star formation or color vs. $\Sigma_1$ or $\Delta\Sigma_1$ (\citealt{Fang13, Barro17, Lee2018, Luo20, Guo2021}). Following the trend in KM means, the H$_2$ fraction exhibits minor variation until just before $\Delta\Sigma_1 = 0$, at which point the mean fraction decreases by an order of magnitude over an interval of $\Delta\Sigma_1\approx 0.4$ dex. This trend is also seen in the overlap sample (Figure \ref{fig:delSigma1_sSFR}).

We observe that the trend between $\Delta\Sigma_1$ and $f_\mathrm{H2}$ is steeper than the relationship between $\Delta\Sigma_1$ and $f_\mathrm{HI}$ in the vicinity of the $\Delta\Sigma_1 = 0$ point.
To quantify this, we apply the \verb|LINMIX_ERR| method to fit a line to the data in the regime of $-0.2 < \log\Delta\Sigma_1 < 0.2$, where it is still possible to fit a line, within error, to the KM means. We find that the slope between $\log f_\mathrm{H2}$ and $\log\Delta\Sigma_1$ is $-2.57 \pm 0.29$, compared to a slope of $-2.16 \pm 0.21$ between $\log f_\mathrm{HI}$ and $\log\Delta\Sigma_1$. For the overlap sample, we obtain results consistent with these values, so our finding that the slope observed for $\log f_\mathrm{H2}$ vs. $\log\Delta\Sigma_1$ is steeper than the $\log f_\mathrm{HI}$ vs. $\log\Delta\Sigma_1$ slope is robust against possible sampling bias due to the different samples. Our observation that the $\Delta\Sigma_1-f_\mathrm{H2}$ trend is steeper than its $f_\mathrm{HI}$ counterpart around the $\Delta\Sigma_1 = 0$ point suggests that the molecular gas fraction may be more sensitive to changes in $\Sigma_1$ at fixed mass, particularly where galaxies cross the $\Sigma_1 - M_\ast$ structural valley. 

In agreement with \cite{Luo20}, we find that there is a critical point marked by $\Delta\Sigma_1 = 0$ after which point star formation is generally suppressed. This is illustrated in Figure \ref{fig:delSigma1_sSFR}, where the shading of points indicates that sSFR declines with increasing $\Delta\Sigma_1$ and with decreasing $f_\mathrm{HI}$ and $f_\mathrm{H2}$. Galaxies with lower sSFR are found preferentially where $\Delta\Sigma_1 > 0$. 

To quantify this, we employ the star-forming criterion of \cite{Woo15} to define the fraction of quenched objects as a function of $\Delta\Sigma_1$ (Figure \ref{fig:delSigma1_sSFR}). Among galaxies with $\Delta\Sigma_1 > 0$, we find that $65.6\%$ of our H{\small I} sample and $62.3\%$ of our H$_2$ sample are quiescent. In contrast, only $7.2\%$ and $5.6\%$ of galaxies with $\Delta\Sigma_1 < 0$ are quiescent in our H{\small I} and H$_2$ samples, respectively. Consequently, we contend that $\Delta\Sigma_1 = 0$ approximately marks a transition from star-forming to quiescent galaxies.

\subsection{AGN Activity}\label{subsec:results_agn}

In Figure \ref{fig:delSigma1_AGN}, we show the distribution of Seyfert and LINER host candidates in the plane of cold gas fraction vs. $\Delta\Sigma_1$. For galaxies that have $\Delta\Sigma_1 > 0$, we find that $60.2\%$ of the H{\small I} sample and $58.5\%$ in the H$_2$ sample are classified as AGN host candidates (either Seyfert or LINER), while just$16.5\%$ and $12.7\%$ of all galaxies with $\Delta\Sigma_1 < 0$ are AGN candidates in our H{\small I} and H$_2$ samples, respectively. 

While AGN are overrepresented in our sample compared to the \cite{Luo20} sample (Section \ref{sec:BPT_diagram}), the general trend is the same between their sample and ours; there is an increase in AGN activity for $\Delta\Sigma_1 > 0$, and little AGN activity for $\Delta\Sigma_1 < 0$. Within the \cite{Luo20} sample of galaxies -- after we impose the constraints that $M_\ast > 10^{10} M_\odot$ and $0.02 \leq z \leq 0.05$ to be consistent with our samples -- approximately $50\%$ of the galaxies located above the $\Sigma_1-M_\ast$ structural valley ($\Delta\Sigma_1 > 0$) are AGN candidates, compared to $\approx 12\%$ of galaxies below. This suggests that AGN candidates are possibly overrepresented in our samples for $\Delta\Sigma_1 > 0$. Regardless, in both our sample and the parent sample, we can conclude that the threshold $\Delta\Sigma_1 > 0$ is associated with a greater incidence of AGN candidates relative to the region $\Delta\Sigma_1 < 0$. 

To further visualize the connection between AGN activity, inner stellar surface mass density and gas depletion, we separately track the relative proportion of Seyfert/LINER candidates in bins of $\Delta\Sigma_1$ in panels (e) and (f) of Figure \ref{fig:delSigma1_AGN}. Uncertainties in the binned fractions are obtained via bootstrapping.

We investigate Seyfert and LINER trends separately due to the fact that many LINERs may be produced by ionization due to post-AGB stars rather than AGN (e.g. \citealt{Singh_2013}). To supplement the BPT diagram, previous works have imposed the additional constraint that galaxies have an H$\alpha$ equivalent width EW$_\mathrm{H\alpha} > 3$ \AA \ to be considered genuine AGN (\citealt{WHAN_2010, WHAN_2011}). We compute EW$_\mathrm{H\alpha}$ values for the LINERs in our H{\small I} and H$_2$ samples from the MPA/JHU DR7 catalog. We find that only $\sim 10\%$ and $\sim 8 \%$ of the LINERs have EW$_\mathrm{H\alpha} > 3$ \AA \ in our H{\small I} and H$_2$ samples, respectively. For this reason, we caution the reader against interpreting our LINER sample as genuine AGN, but rather as weak AGN \textit{candidates} or retired galaxies.

We find that Seyfert activity peaks at $\Delta\Sigma_1 \approx 0.2$ dex, and the abundance of LINER hosts increases greatly between $\Delta\Sigma_1\approx0$ and $0.2$ dex (Figure \ref{fig:delSigma1_AGN}). Overall, $>60\%$ of galaxies that have $\Delta\Sigma_1 > 0.2$ dex are AGN candidates, which indicates that elevated central surface mass density at fixed mass is associated with a higher likelihood of AGN activity.

\section{Discussion} \label{sec:discussion} 

\subsection{The Relationship between \texorpdfstring{$\Sigma_1$}{Sigma1} and Cold Gas} \label{discussionsigma1}

Our findings suggest that $\Sigma_1$ is anti-correlated with both $f_\mathrm{HI}$ and $f_\mathrm{H2}$ for $\Sigma_1 > 10^9 \ M_\odot \ \mathrm{kpc}^{-2}$. However, the trends between the mean atomic and molecular gas fractions and $\Sigma_1$, shown by the red points in panels (c) and (f) of Figure \ref{fig:gas_fraction}, are notably different, mirroring the relationship between gas fraction and $\mu_\ast$. For $\Sigma_1 < 10^9 \ M_\odot \ \mathrm{kpc}^{-2}$, the trend between $f_\mathrm{H2}$ and $\Sigma_1$ is flat, but there is a quantifiable decline of approximately $1$ dex in $f_\mathrm{H2}$ for $10^9 < \Sigma_1 / (M_\odot \ \mathrm{kpc}^{-2}) < 10^{10}$. The relationship between $f_\mathrm{HI}$ and $\Sigma_1$ does not appear to adhere to an `elbow' shape which would suggest nonlinear behavior. Although the correlation is weak for $\Sigma_1 < 10^9 \ M_\odot \ \mathrm{kpc}^{-2}$, the mean atomic gas fraction generally follows a decreasing trend across the full range of $\Sigma_1$, which is not the case for $f_\mathrm{H2}$. Because $f_\mathrm{H2}$ appears to change only at a particular threshold in $\Sigma_1$, we postulate that the molecular gas fraction may be more directly affected by the growth of the central bulge than the atomic gas fraction.

As the detected sample consists predominantly of main-sequence galaxies, this observation could be explained by appealing to the typical galactic distribution of cold gas in massive disk galaxies. H{\small I} is accreted into the outskirts of galaxies via gas infall from the intergalactic medium or cosmic filaments, after which point it can replenish the inner parts of the galaxy and fuel star formation via conversion into the molecular phase (see e.g. \cite{Sancisi2008} for a review of cold gas accretion). As a result, the centers of nearby spiral galaxies are dominated by H$_2$ gas, while the outskirts are dominated by H{\small I} gas (\citealt{Young91, Bigiel2008, Leroy2008}). Since $\Sigma_1$ probes the central regions of galaxies, we expect $f_\mathrm{H2}$ to exhibit greater sensitivity to $\Sigma_1$ than $f_\mathrm{HI}$. This is further supported by the more direct connection between $\Delta\Sigma_1$ and $f_\mathrm{H2}$ compared to $f_\mathrm{HI}$, which is discussed in depth in the next subsection. However, it should be noted that the mild `elbow' shape observed in the plane of $f_\mathrm{H2}$ vs. $\Sigma_1$ could be a byproduct of the tighter correlation between $f_\mathrm{H2}$ and SFR, and of the nonlinear behavior which has been reported between $\Sigma_1$ and sSFR (e.g. \citealt{Barro13}). 

\subsection{\texorpdfstring{$\Delta\Sigma_1$}{DeltaSigma1} as an Indictor of Evolution Towards Quenching}

Previous literature has proposed that galaxies evolve towards higher $\Sigma_1$ and that galaxies quench at some mass-dependent threshold (\citealt{Fang13, vanDokkum2015, Barro17}). \cite{Luo20} contended that $\Delta\Sigma_1 =0$ is the demarcation between pseudo-bulges, which lie to the left of $\Delta\Sigma_1 = 0$, and classical bulges, which lie to the right. They observed a ``bimodality" in the distribution of $\Delta\Sigma_1$ vs. global sSFR, whereby galaxies to the left of $\Delta\Sigma_1 = 0$ have systematically greater values of sSFR, while galaxies to the right have lower values, but with some range in sSFR. \cite{Luo20} raised the possibility that some galaxies evolve from pseudo- to classical-type bulges, or from low to high $\Delta\Sigma_1$, and as such they offered an explanation for the greater range of sSFR values to the right of $\Delta\Sigma_1 = 0$ as the result of differences in structure and stellar populations during evolution towards quenching. 

We observe a decrease in both $f_\mathrm{H2}$ and $f_\mathrm{HI}$ across $\Delta\Sigma_1 = 0$, and a corresponding decrease in sSFR (Figure \ref{fig:delSigma1_sSFR}). While $f_\mathrm{HI}$ exhibits an approximately linear decline with $\Delta\Sigma_1$, $f_\mathrm{H2}$ decreases nonlinearly with $\Delta \Sigma_1$, particularly across the $\Delta\Sigma_1$ zero point as discussed in Section \ref{results_delSigma1}. This makes sense in light of the fact that the central gas reservoir is dominated by molecular gas, as discussed in Section \ref{discussionsigma1}. Additionally, there is evidence to suggest that H$_2$ content is more tightly correlated with galaxy quiescence than H{\small I}. \cite{Zhang2019} observed evidence for significant H{\small I} gas reservoirs in nearby, massive quiescent galaxies which were classified as centrals with disk morphology. A significant proportion of their sample had H{\small I} detections but were generally depleted in H$_2$ gas.

In accordance with \cite{Luo20}, we find that $\Delta\Sigma_1$, like $\Sigma_1$, is indicative of quenching, and furthermore, that $\Delta\Sigma_1$ may be indicative of \textit{evolution} towards quenching. It is tempting to consider that the mean trend of $\Delta\Sigma_1$ vs. $f_\mathrm{H2}$ in Figures \ref{fig:delSigma1_sSFR} and \ref{fig:delSigma1_AGN} constitutes an evolutionary track; in the remainder of this subsection, we show that this is plausible. 

We observe that sSFR systematically decreases with $f_\mathrm{H2}$ and $\Delta\Sigma_1$, and that where the mean trend line crosses the $\Delta\Sigma_1 = 0$ point in Figure \ref{fig:delSigma1_sSFR}, $\mathrm{sSFR} \sim 10^{-11} \mathrm{yr}^{-1}$. The delineation between star-forming and quiescent galaxies as shown in Figure \ref{fig:contour} corresponds to $\mathrm{sSFR} \sim 10^{-11} \mathrm{yr}^{-1}$; hence, $\Delta\Sigma_1=0$ also approximately marks the quenching point along the mean trend line.

\cite{Fang13} contend that the key evolutionary driver of quenching for massive, central galaxies in the local Universe is the build-up of central mass surface density at roughly fixed global stellar mass. An evolutionary track within the plane of $\Delta\Sigma_1$ vs. $f_\mathrm{H2}$ must therefore generally occur in the direction of increasing $\Delta\Sigma_1$, at least before the quenching point. While there are processes which can reduce $\Sigma_1$, the effects are generally minor and occur after quenching (\citealt{vanDokkum2015, Barro17}). Together with the fact that sSFR should decline as galaxies quench, then galaxies on this evolutionary track must also be declining in sSFR. The mean trend of Figure \ref{fig:delSigma1_sSFR} fulfills these criteria. Within this interpretation, then on average, H$_2$ gas is depleted more quickly at fixed mass than H{\small I} gas within the vicinity of the quenching point, $-0.2 \ \mathrm{dex} \lesssim \log\Delta\Sigma_1 \lesssim 0.2 \ \mathrm{dex}$. This implies that molecular gas is depleted faster than atomic gas as galaxies evolve towards quenching, as marked by central bulge growth. \citealt{Popping2015} shows that molecular gas evolves much more rapidly than atomic gas over cosmic time, with $f_\mathrm{H2}$ declining strongly while H{\small I} remains nearly constant at fixed stellar mass. In our nearby-galaxy sample, we find the same trend in the rate of change: $f_\mathrm{H2}$ drops steeply across the $\Delta\Sigma_1 = 0$ transition, whereas $f_\mathrm{HI}$ decreases more gradually.

We caution against interpreting the mean trend as a \emph{single} evolutionary track; it is possible that there are multiple evolutionary tracks within the $f_\mathrm{H2}-\Delta\Sigma_1$ plane, a point which we refer to in the next subsection. Additionally, there may be exceptions to our assumptions of galaxies building up their central bulges at fixed mass. Deviations from the fixed-mass, bulge-growth picture could be an additional source of scatter in our data. 

\cite{Ma2022} employed cosmological simulations to examine how $\Delta\Sigma_1$ correlates with offsets in cold gas mass—specifically, $\Delta \log M_\mathrm{H{\small I}MS}$ and $\Delta \log M_\mathrm{H_2MS}$—where each is defined as the deviation in either $M_\mathrm{HI}$ or $M_\mathrm{H_2}$ from the corresponding $M_\mathrm{gas}$–$M_\ast$ ``main sequence." In their work, this main sequence is constructed as the median value of $\log M_\mathrm{gas}$ in bins of $\log M_\ast$ for star-forming galaxies (SFGs); similarly, $\Delta\Sigma_1$ is defined as the deviation from the median $\Sigma_1$ in bins of $M_\ast$ for SFGs. Our results reproduce the general trends of their Figure 11, although we find a steeper correlation between H$_2$ content and $\Delta\Sigma_1$ than reported by \cite{Ma2022}. While $f_\mathrm{H2}$ declines by $\sim 1$ dex between $\Delta\Sigma_1 \approx -0.2$ and $0.2$ dex in our sample, \cite{Ma2022} find that $\log M_\mathrm{H_2 MS}$ decreases by only $\sim 0.25$ dex in this same range in $\Delta\Sigma_1$. We also observe a flatter trend between H{\small I} gas fraction and $\Delta\Sigma_1$ than they report. We attribute these discrepancies partly to our different definition of $\Delta\Sigma_1$ and, to our choice of using cold gas fractions (i.e., $M_\mathrm{H_2}/M_\ast$) rather than a cold gas main sequence. Our primary goal is to assess $\Delta\Sigma_1$ as an indicator of quenching, so employing the definition from \cite{Luo20} -- which is independent of any star-forming or quenched designation -- offers a potentially less biased parameterization for tracking the depletion of cold gas during quenching. Regardless of the technical differences, our work provides observational evidence supporting the conclusion of \cite{Ma2022}.

To summarize, our results in Figure \ref{fig:delSigma1_sSFR} suggest that the growth of the inner stellar mass surface density at fixed mass is more closely linked to a depletion in H$_2$ gas content rather than H{\small I} content, as indicated by the greater decrease in gas fraction in the vicinity of $\Delta\Sigma_1 = 0$.  This trend is also coincident with a decrease in sSFR. The primary implication of this trend, if interpreted as an evolutionary one, is that massive galaxy quenching is accompanied predominantly by the removal or depletion of molecular gas rather than atomic gas as the central bulge grows.

However, we cannot rule out the possibility that the apparent differences in how $f_\mathrm{HI}$ and $f_\mathrm{H2}$ depend on $\Delta\Sigma_1$ are merely byproducts of other correlations -- namely the bimodal relationship between $\Delta\Sigma_1$ and sSFR, and the different dependencies of molecular and atomic gas fractions on sSFR. In particular, $f_\mathrm{H2}$ correlates more strongly with sSFR than does $f_\mathrm{HI}$ (e.g. \citealt{Saintonge22}). Moreover, because H$_2$ is typically more concentrated in galaxy centers, the differing trends observed between $f_\mathrm{HI}$ and $f_\mathrm{H2}$ with $\Delta\Sigma_1$ could simply reflect quenching mechanisms associated with central bulge growth and with the resulting reduction in global sSFR, rather than physical differences in H{\small I} vs. H$_2$ gas depletion. In any case, our results support that the interplay of these three parameters is important for the quenching of massive galaxies at fixed mass.

\subsection{The Connection to AGN Activity}
As seen in Figure \ref{fig:delSigma1_AGN}, we observe a fairly significant decrease in $f_\mathrm{H2}$ and a complementary increase in the number of AGN host candidates -- both Seyfert and LINER -- across $\Delta\Sigma_1 = 0$. Superficially, this fits into the picture of inside-out quenching, possibly aided by AGN feedback. Combined with the fact that H$_2$ gas is concentrated more in the centers of galaxies than {H\small I} gas, our findings are suggestive of a model by which the gas available for star formation is either expelled predominantly from the inner regions of the galaxy, or prevented from cooling or accumulating further near the galactic center by some other means; gas in the outskirts of the galaxy is not as affected. The coincident rise of potential AGN activity, gas depletion and central bulge growth could suggest that AGN feedback is at least in part responsible for the decrease of cold gas in the center regions of galaxies, but we cannot ascertain this causation from the nature of our analysis. At minimum, our results show that AGN activity is associated with lower cold gas fractions.

Nonetheless, we can postulate as to the relevance and modes of AGN feedback in our sample from an investigation of trends in Seyfert and LINER abundances with gas fraction and $\Delta\Sigma_1$. It is interesting to note that some of the greatest changes in observable properties -- including Seyfert and LINER abundances -- in our sample occur between $\Delta\Sigma_1 \approx 0$ dex and $0.2$ dex (Figures \ref{fig:delSigma1_sSFR} and \ref{fig:delSigma1_AGN}). If galaxies begin to cross the green valley at $\Delta\Sigma_1 = 0$, marking the point at which galaxies begin to quench, then it would seem that most galaxies ($>70$\%) are quenched by $\Delta\Sigma_1 \approx 0.2$ dex and mostly depleted in molecular and atomic gas (Figure \ref{fig:delSigma1_sSFR}).

The greatest abundance of Seyfert candidates relative to the overall population occurs near $\Delta\Sigma_1 \approx 0.2$ dex (Figure \ref{fig:delSigma1_AGN}). It is well known that Seyfert nuclei tend to be hosted by bulge-dominated, early-type spiral galaxies (e.g. \citealt{Moles_1995}), so it is unsurprising that Seyfert hosts tend to have elevated $\Delta\Sigma_1$. However, while there is sufficient evidence for the effects of Seyfert AGN feedback on the innermost regions of galaxies, studies of the global cold gas contents of local Seyfert host galaxies have yielded mixed results. High-mass Seyfert hosts may not exhibit differences in their galactic-scale H$_2$ fractions compared to SFGs of the same mass (\citealt{Salvestrini_2022}), and H{\small I} observations of Seyfert hosts suggest that some may actually be enriched in their H{\small I} gas fractions (e.g. \citealt{Hunt_1999}). In Figure \ref{fig:delSigma1_AGN}, we find that Seyfert hosts exhibit a range of H{\small I} and H$_2$ gas fractions, reflecting the complex relationship between Seyfert nuclei and the gas reservoirs of their host galaxies. It is also challenging to infer the influence of Seyfert AGN feedback on population-level trends, since only 20\% of the population at $\log\Delta\Sigma_1 \approx 0.2$ dex is categorized as being in a Seyfert phase (Figure \ref{fig:delSigma1_AGN}).

It is well known that there is a greater abundance of LINERs in the nearby Universe, which is reflected in our Figure \ref{fig:delSigma1_AGN}. While LINER-like emission may trace older stellar populations rather than AGN activity (e.g. \citealt{Singh_2013}), still other studies contend that many LINERs in the nearby Universe do represent genuine, low-luminosity AGN, possibly even fueling outflows (e.g. \citealt{Marquez_2017}). Regardless of the source of the emission, some LINERs may be responsible for maintaining quiescence in nearby galaxies via ``maintenance mode feedback", as evidenced by kinematic disturbances in ionized gas (e.g. \citealt{Gatto2024}). The galaxies in our sample with LINER-like emission-line flux ratios are predominantly quenched, gas-poor systems. While approximately 90\% of these galaxies have EW$_\mathrm{H\alpha} < 3$ \AA, which is consistent with photoionization by older stars, this does not rule out the possibility that many of these galaxies were once powered by AGN and are now retired.

We additionally consider if our BPT classification could have missed obscured AGN in the star-forming and composite regions of the diagram. Although it is possible for AGN to be obscured by dust on galactic scales (\citealt{Hickox2018}), our sample selection minimizes this concern. The \cite{Luo20} catalog excludes edge-on systems, mitigating the impact of dust attenuation in the host galaxy that can suppress optical AGN signatures (\citealt{Hickox2018}). Obscured AGN are also more commonly associated with gas-rich mergers, where inflows can fuel both star formation and nuclear activity behind substantial columns of dust (\citealt{Hopkins2008}). To confirm that our sample does not include mergers, we cross-matched our sample with Galaxy Zoo classifications (\citealt{Lintott11, Willett13}) and found no objects explicitly defined as a merger. Given the absence of mergers and our selection against edge-on geometries, it is unlikely that we are missing a population of heavily obscured AGN.

Our result that fixed-mass central bulge growth is connected to possible AGN activity and molecular gas depletion, rather than atomic gas depletion, is consistent with other studies of the impact of AGN feedback on massive galaxies in the nearby Universe. It has been suggested that AGN feedback is not likely a primary cause of global HI depletion in nearby galaxies (\citealt{Fabello_2011, Gereb_2015}). There is evidence for galactic-scale outflows of molecular gas driven by AGN feedback (e.g. \citealt{Cicone14}), and \cite{Fluetsch21} observed that for a small sample of local, ultra-luminous infrared galaxies, a higher fraction of AGN-driven outflows consist of molecular rather than neutral atomic gas.

We now return to the discussion of scatter in Figures \ref{fig:delSigma1_sSFR} and \ref{fig:delSigma1_AGN}. While our data suggests that AGN activity and bulge growth are closely related to massive central galaxy quenching, there may be other potential quenching mechanisms at play. $\Delta\Sigma_1$ can only provide direct information about the innermost region of the galaxy, and so the role of halo or environmental effects are not captured in our work. Accounting for halo or environment properties would illuminate other possible quenching mechanisms and could potentially explain the observed variability in the $\Delta\Sigma_1-f_\mathrm{H2}$ plane beyond measurement uncertainties. In addition to gas removal from the disk, recent accretion events could also contribute to the variation.

Our sample was predominantly limited to galaxies with stellar masses $M_\ast > 10^{10} M_\odot$. Future work may test the relationship between $\Delta\Sigma_1$ and cold gas fraction at stellar masses down to $M_\ast \sim 10^9 M_\odot$ to encapsulate the full range of main-sequence galaxies, which are generally supposed to quench in an inside-out fashion. 

Additionally, with integral field spectroscopy (IFS), it is possible to compare the $\Delta\Sigma_1$ values for a galaxy with the distribution and kinematics of its gas reservoir. Such an analysis could offer insight into the mechanisms through which the growth of $\Sigma_1$ produces changes to the surrounding gas reservoir. There is evidence for centrally suppressed sSFR for galaxies just below the star-forming main sequence (\citealt{Tacchella18}). Recent IFS studies with the Atacama Large Millimeter Array have supplemented previous spatially resolved studies with sub-kpc evidence for the imprint of AGN activity on molecular gas reservoirs (e.g. \citealt{Garcia-Burillo2024, Holden2024}). Investigations of this nature could test the strength of $\Delta\Sigma_1$ as a more readily available tracer for quenching and AGN activity, which can be applied to large samples for statistical analyses.

\section{Conclusion} \label{sec:conclusion}

We have investigated the relationship between cold gas content and $\Sigma_1$ in nearby galaxies, using observations from the xGASS and xCOLDGASS surveys. We cross-matched HI and H$_2$ data with $\Sigma_1$ values measured by \cite{Luo20}, resulting in a sample of 341 galaxies with HI data and 201 with H$_2$ data. By analyzing cold gas fractions as functions of $\Sigma_1$ and the relative parameter $\Delta\Sigma_1$, we assessed how inner galactic structure is connected to the decline in global cold gas reservoirs as galaxies evolve towards quenching. Furthermore, we explored the connection between AGN activity, $\Delta\Sigma_1$, and gas fraction, offering insight into quenching mechanisms in the local Universe. We obtained the following results:

\begin{itemize}
    \item We find comparable correlations between both the H{\small I} and H$_2$ cold gas fractions and $\Sigma_1$ in nearby galaxies in the regime $\Sigma_1 \gtrsim 10^9 \ M_\odot \ \mathrm{kpc}^{-2}$. These trends are similar in strength to previously established relationships between cold gas fractions vs.~$M_{\ast}$ and $\mu_{\ast}$ (e.g. \citealt{Saintonge22}).

    \item The trend between $f_\mathrm{H2}$ and $\Delta\Sigma_1$ is essentially flat for galaxies below $\Delta\Sigma_1\approx0$ then steepens above $\Delta\Sigma_1\approx0$. This ``elbow'' shape is not seen in the H{\small I} gas fraction, which suggests a more direct link between \emph{molecular} gas depletion and mass buildup in the bulge. 

    \item We observe two different regimes in the plane of $f_\mathrm{H2}$ vs. $\Delta\Sigma_1$. Galaxies with $ \Delta\Sigma_1 < 0$ dex tend to be gas-rich and star-forming, while those with $\Delta\Sigma_1 > 0$ dex tend to be gas-poor and quenched. We observe a steep decline in $f_\mathrm{H2}$ between $\Delta\Sigma_1 \approx 0$ and $0.2$ dex. This threshold region marks the transition to quiescence and is accompanied by a substantial rise in AGN candidates. We propose that galaxies may evolve across this plane, undergoing a decline in their molecular gas fractions as their central bulges grow, on a path consistent with inside-out quenching and a decrease in global sSFR.

    \item We observe a greater abundance of AGN host candidates for $\Delta\Sigma_1 > 0$. This is consistent with the connection between AGN activity and central bulge growth, as well as the possible role of AGN feedback in gas depletion and quenching. For $\Delta\Sigma_1 \gtrsim 0.2$ dex, more than $70\%$ of galaxies are quenched, over $60\%$ of galaxies are AGN host candidates, and the downturn in both the H{\small I} and H$_2$ gas fractions levels off.
    
    \item Our findings support an inside-out quenching model in which molecular gas is preferentially depleted from the central regions of galaxies, or prevented from accumulating further, as the bulge grows, with atomic gas reservoirs being less directly impacted at fixed galaxy mass.
\end{itemize}

Prior research has shown that AGN hosts, including Seyferts and LINERs, often exhibit central deficits in molecular gas and signs of turbulence or outflows that can suppress star formation. However, while our data suggest an association between elevated $\Delta\Sigma_1$, reduced $f_\mathrm{H2}$, and increased AGN activity, the evidence does not confirm a direct causal relationship. Future work may involve detailed analyses of galaxies classified as Seyfert or LINER candidates that have both $f_\mathrm{HI}$ and $f_\mathrm{H2}$ measurements. This would allow us to probe whether these AGN candidates actively drive gas depletion or simply arise in already gas-poor systems. Additionally, measuring gas outflow rates can help determine whether AGN are capable of removing gas at levels sufficient to halt star formation. 

This work underscores the utility of $\Sigma_1$ and $\Delta\Sigma_1$ as observational tools for tracing galaxy evolution and quenching. As larger samples and higher-resolution gas data become available, leveraging these structural parameters can help untangle the complex interplay between central bulge growth, gas depletion, and feedback mechanisms in the life-cycle of galaxies.

\begin{acknowledgments}
E. E. Shread and T. J. Weiss contributed equally to this manuscript and may both refer to this as a first-author publication. This work was funded by the Orange Coast College Foundation. We thank the anonymous referee for their insightful comments that improved this manuscript. We thank Sandra Faber for her advice on data analysis and future steps, Renata Cioczek-Georges for her insight on statistical analysis of the data, and Jessica Asbell, Yicheng Guo, and Matt Malkan for their helpful comments and insights. Thanks to Paulo Barchi, Reinaldo de Carvalho, Igor Kolesnikov, and the CyMorph team for sharing their morphological data during the early stages of this work.
\end{acknowledgments}

\software{ SciPy (\citealt{SciPy20}),
astropy (\citealt{astropy13, astropy18, astropy22}), 
pandas (\citealt{pandas20}),
Matplotlib (\citealt{Hunter07}),
seaborn (\citealt{Waskom21}),
numpy (\citealt{Harris20}),
lifelines (\citealt{davidson_pilon_2024_14007206})}

\vspace{5mm}
\facilities{Arecibo, GALEX, WISE, IRAM-30m, APEX, SDSS, APO:2.5m}

\bibliography{Bibliography}{}

@PREAMBLE{
 "\providecommand{\noopsort}[1]{}"  
}

@ARTICLE{Young89,
   author       = "Judith S. Young and Patricia M. Knezek",
   year         = "1989",
   journal      = "ApJ",
   volume       = "347",
   pages        = "L55-L58",
   doi          = {10.1086/185606}
}

@ARTICLE{Young91,
   author       = "J. S. Young and N. Z. Scoville", 
   year         = "1991", 
   journal      = "ARAA", 
   volume       = "29", 
   pages        = "581-625",
   doi         = {10.1146/annurev.aa.29.090191.003053}
}

@ARTICLE{Roberts94,
    author = "Morton S. Roberts and Martha B. Haynes", 
    year = "1994", 
    journal = "ARAA", 
    volume = "32", 
    pages = "115-152",
    url ={https://doi.org/10.1146/annurev.aa.32.090194.000555},
    doi = {10.1146/annurev.aa.32.090194.000555}
}

@ARTICLE{Saintonge17,
    author      = "A. Saintonge and B. Catinella and L. J. Tacconi and others",
    year        = "2017",
    journal     = "ApJS",
    volume      = "233",
    pages       = "22-41",
    doi          = {10.3847/1538-4365/aa97e0}
}

@ARTICLE{Saintonge22,
    author      = "A. Saintonge and B. Catinella",
    year        = "2022",
    journal     = "ARAA",
    volume      = "60",
    pages       = "319-361",
    doi         ={10.1146/annurev-astro-021022-043545}
}

@ARTICLE{Catinella10,
    author      = "Barbara Catinella and David Schiminovich and Guinevere Kauffmann and others",
    year        = "2010",
    journal     = "MNRAS",
    volume      = "403",
    pages       = "683-708",
    doi         ={10.1111/j.1365-2966.2009.16180.x}
}

@ARTICLE{Brown15,
    author      = "Toby Brown and Barbara Catinella and Luca Cortese and others",
    year        = "2015",
    journal     = "MNRAS",
    volume      = "452",
    pages       = "2479–2489",
    doi         ={10.1093/mnras/stv1311}
}

@ARTICLE{Hunt20,
    author      = "L. K. Hunt and C. Tortora and M. Ginolfi and others",
    year        = "2020",
    journal     = "AAP",
    volume      = "643",
    pages       = "A180",
    doi         ={10.1051/0004-6361/202039021}
}

@ARTICLE{Sage93,
    author      = "L. J. Sage",
    year        = "1993",
    journal     = "A\&A",
    volume      = "100",
    pages       = "537-569",
}

@ARTICLE{Davis22,
    author      = "Timothy A. Davis and Jindra Gensior and Martin Bureau and others",
    year        = "2022",
    journal     = "MNRAS",
    volume      = "512",
    pages       = "1522-1540",
    doi         ={10.1093/mnras/stac600}
}

@ARTICLE{Fang13,
    author      = "Jerome J. Fang and S.M. Faber and David C. Koo and others",
    year        = "2013",
    journal     = "ApJ",
    volume      = "776",
    pages       = "63-81",
    doi         ={10.1088/0004-637X/776/1/63 }
}

@ARTICLE{Luo20,
    author      = "Yifei Luo and S.M. Faber and Aldo Rodri\'guez-Puebla and others",
    year        = "2020",
    journal     = "MNRAS",
    volume      = "493",
    pages       = "1686-1709",
    doi         ={10.1093/mnras/staa328}
}

@ARTICLE{Woo15,
    author      = "Joanna Woo and S.M. Faber and Avishai Dekel  and others",
    year        = "2015",
    journal     = "MNRAS",
    volume      = "448",
    pages       = "237-251",
    doi         ={10.1093/mnras/stu2755}
}

@ARTICLE{Lin20,
    author      = "Lin Lin and S.M. Faber and David C. Koo and others",
    year        = "2020",
    journal     = "ApJ",
    volume      = "899",
    pages       = "93-109",
    doi         ={10.3847/1538-4357/aba755}
}

@ARTICLE{Boselli14,
    author      = "A. Boselli and L. Cortese and M. Boquien and others",
    year        = "2014",
    journal     = "A\&A",
    volume      = "564",
    pages       = "A66-A84",
    doi         ={10.1051/0004-6361/201322312}
}

@ARTICLE{Lintott11,
    author      = "Chris J. Lintott and Kevin Schawinski and Steven Bamford and others",
    year        = "2011",
    journal     = "MNRAS",
    volume      = "410",
    pages       = "166-178",
    doi         ={10.1111/j.1365-2966.2010.17432.x}
}

@ARTICLE{Namiki21,
    author      = "Shgieru V. Namiki and Yusei Koyama and Shuhei Koyama and others",
    year        = "2021",
    journal     = "ApJ",
    volume      = "918",
    pages       = "68-81",
    doi         ={10.3847/1538-4357/abfe08}
}

@ARTICLE{Zhou18,
    author      = "Zhimin Zhou and Hong Wu and Xu Zhou and others",
    year        = "2018",
    journal     = "PASP",
    volume      = "130",
    pages       = "1-25",
    doi         ={10.1088/1538-3873/aad407}
}

@ARTICLE{Yesuf19,
    author      = "Hassan M. Yesuf and Luis C. Ho",
    year        = "2019",
    journal     = "ApJ",
    volume      = "884",
    pages       = "177-196",
    doi         ={10.3847/1538-4357/ab4202}
}

@ARTICLE{Kauffmann12,
    author      = "Guinevere Kauffmann and Cheng Li and Jian Fu and others",
    year        = "2012",
    journal     = "MNRAS",
    volume      = "442",
    pages       = "997-1006",
    doi         ={10.1111/j.1365-2966.2012.20672.x}
}

@ARTICLE{Huang12,
    author      = "Shan Huang and Martha P. Haynes and Riccardo Giovanelli and others",
    year        = "2012",
    journal     = "ApJ",
    volume      = "756",
    pages       = "113-139",
    doi         ={10.1088/0004-637X/756/2/113}
}

@ARTICLE{Brinchmann04,
    author      = "J. Brinchmann and S. Charlot and S.D.M White and others",
    year        = "2004",
    journal     = "MNRAS",
    volume      = "351",
    pages       = "1151-1179",
    doi         ={10.1111/j.1365-2966.2004.07881.x}
}

@ARTICLE{Catinella18,
    author      = "Barbara Catinella and Amelia Saintonge and Steven Janowiecki and others",
    year        = "2018",
    journal     = "MNRAS",
    volume      = "476",
    pages       = "875-895",
    doi         = {10.1093/mnras/sty089}
}

@ARTICLE{Conselice14,
    author      = "Christopher J. Conselice",
    year        = "2014",
    journal     = "ARAA",
    volume      = "476",
    pages       = "875-895",
    doi         ={10.1146/annurev-astro-081913-040037}
}

@ARTICLE{Willett13,
    author      = "Kyle W. Willett and Chris J. Lintott and  Steven P. Bamford and others",
    year        = "2013",
    journal     = "MNRAS",
    volume      = "435",
    pages       = "2835-2860",
    doi         ={10.1093/mnras/stt1458}
}

@ARTICLE{Abazajian09,
    author      = "Kevork N. Abazajian and Jennifer K. Adelman-McCarthy and  Marcel A. Agueros and others",
    year        = "2009",
    journal     = "ApJS",
    volume      = "182",
    pages       = "543-558",
    doi         = {10.1088/0067-0049/182/2/543}
}

@ARTICLE{Catinella13,
    author = {Catinella, Barbara and Schiminovich, David and Cortese, Luca and Fabello, Silvia and Hummels, Cameron B. and Moran, Sean M. and Lemonias, Jenna J. and Cooper, Andrew P. and Wu, Ronin and Heckman, Timothy M. and Wang, Jing},
    title = {The GALEX Arecibo SDSS Survey – VIII. Final data release. The effect of group environment on the gas content of massive galaxies},
    journal = {MNRAS},
    volume = {436},
    number = {1},
    pages = {34-70},
    year = {2013},
    month = {09},
    issn = {0035-8711},
    doi = {10.1093/mnras/stt1417},
    url = {https://doi.org/10.1093/mnras/stt1417},
    eprint = {https://academic.oup.com/mnras/article-pdf/436/1/34/18751815/stt1417.pdf},
}

@ARTICLE{Cheung12,
    author      = "Edmond Cheung and S.M. Faber and  David C. Koo and others",
    year        = "2012",
    journal     = "ApJS",
    volume      = "760",
    pages       = "131-156",
    doi         = {10.1088/0004-637X/760/2/131}
}

@ARTICLE{Barro17,
    author      = "Guillermo Barro and S.M. Faber and  David C. Koo and others",
    year        = "2017",
    journal     = "ApJ",
    volume      = "840",
    pages       = "47-70",
    doi         = {10.3847/1538-4357/aa6b05}
}

@article{Kewley01,
    doi = {10.1086/321545},
    url = {https://dx.doi.org/10.1086/321545},
    year = {2001},
    month = {jul},
    publisher = {},
    volume = {556},
    number = {1},
    pages = {121},
    author = {Kewley, L. J. and Dopita, M. A. and Sutherland, R. S. and Heisler, C. A. and Trevena, J.},
    title = {Theoretical Modeling of Starburst Galaxies},
    journal = {ApJ},
}

@article{Kauffmann03,
    author = {Kauffmann, Guinevere and Heckman, Timothy M. and Tremonti, Christy and Brinchmann, Jarle and Charlot, Stéphane and White, Simon D. M. and Ridgway, Susan E. and Brinkmann, Jon and Fukugita, Masataka and Hall, Patrick B. and Ivezić, Željko and Richards, Gordon T. and Schneider, Donald P.},
    title = {The host galaxies of active galactic nuclei},
    journal = {MNRAS},
    volume = {346},
    number = {4},
    pages = {1055-1077},
    year = {2003},
    month = {12},
    issn = {0035-8711},
    doi = {10.1111/j.1365-2966.2003.07154.x},
    url = {https://doi.org/10.1111/j.1365-2966.2003.07154.x},
    eprint = {https://academic.oup.com/mnras/article-pdf/346/4/1055/18646272/346-4-1055.pdf},
}

@article{Kumari21,
    doi = {10.1051/0004-6361/202140757},
    url = {https://doi.org/10.1051/0004-6361/202140757},
    year = {2021},
    month = {Dec},
    publisher = {},
    volume = {656},
    pages = {A140},
    author = {Nimisha Kumari and Roberto Maiolino and James Trussler and Filippo Mannucci and Giovanni Cresci and Mirko Curti and Alessandro Marconi and Francesco Belfiore},
    title = {The extension of the fundamental metallicity relation beyond the BPT star-forming sequence: Evidence for both gas accretion and starvation},
    journal = {A\&A},
}

@article{Bigiel2008,
doi = {10.1088/0004-6256/136/6/2846},
url = {https://dx.doi.org/10.1088/0004-6256/136/6/2846},
year = {2008},
month = {nov},
publisher = {The American Astronomical Society},
volume = {136},
number = {6},
pages = {2846},
author = {Bigiel, F. and Leroy, A. and Walter, F. and Brinks, E. and de Blok, W. J. G. and Madore, B. and Thornley, M. D.},
title = {THE STAR FORMATION LAW IN NEARBY GALAXIES ON SUB-KPC SCALES},
journal = {AJ},
}

@ARTICLE{Barro13,
    author      = {Guillermo Barro and Sandra M. Faber and Pablo G. P\'erez-Gonz\'alez and others},
    year        = "2013",
    journal     = "ApJ",
    volume      = "765",
    pages       = "104",
    doi         = {10.1088/0004-637X/765/2/104}
}

@article{Woo16,
    author = {Woo, Joanna and Carollo, C. M. and Faber, S. M. and Dekel, Avishai and Tacchella, Sandro},
    title = {Satellite quenching, Galaxy inner density and the halo environment},
    journal = {MNRAS},
    volume = {464},
    number = {1},
    pages = {1077-1094},
    year = {2016},
    month = {09},
    issn = {0035-8711},
    doi = {10.1093/mnras/stw2403},
    url = {https://doi.org/10.1093/mnras/stw2403},
    eprint = {https://academic.oup.com/mnras/article-pdf/464/1/1077/18518449/stw2403.pdf},
}

@article{Cicone14,
    author = {C. Cicone and R. Maiolino and E. Sturm and others},
    journal = {A\&A},
    volume = {562},
    pages = {A21},
    year = {2014},
    doi = {10.1051/0004-6361/201322464},
    url = {https://doi.org/10.1051/0004-6361/201322464},
}

@article{Tacchella18,
doi = {10.3847/1538-4357/aabf8b},
url = {https://dx.doi.org/10.3847/1538-4357/aabf8b},
year = {2018},
month = {may},
publisher = {The American Astronomical Society},
volume = {859},
number = {1},
pages = {56},
author = {Tacchella, S. and Carollo, C. M. and Schreiber, N. M. Förster and Renzini, A. and Dekel, A. and Genzel, R. and Lang, P. and Lilly, S. J. and Mancini, C. and Onodera, M. and Tacconi, L. J. and Wuyts, S. and Zamorani, G.},
title = {Dust Attenuation, Bulge Formation, and Inside-out Quenching of Star Formation in Star-forming Main Sequence Galaxies at z ∼ 2*},
journal = {ApJ},
}

@article{Sancisi2008,
    author = {Renzo Sancisi and Filippo Fraternali and Tom Oosterloo and Thijs van der Hulst},
    title = {Cold gas accretion in galaxies},
    journal = {A\&ARv},
    volume = {15},
    pages = {189-223},
    year = {2008},
    doi = {10.1007/s00159-008-0010-0},
    url =  {https://doi.org/10.1007/s00159-008-0010-0}
}

@article{Popping2015,
    author = {Popping, Gergö and Behroozi, Peter S. and Peeples, Molly S.},
    title = {Evolution of the atomic and molecular gas content of galaxies in dark matter haloes},
    journal = {MNRAS},
    volume = {449},
    number = {1},
    pages = {477-493},
    year = {2015},
    month = {03},
    issn = {0035-8711},
    doi = {10.1093/mnras/stv318},
    url = {https://doi.org/10.1093/mnras/stv318},
    eprint = {https://academic.oup.com/mnras/article-pdf/449/1/477/4147014/stv318.pdf},
}

@article{Lee2018,
doi = {10.3847/1538-4357/aaa40f},
url = {https://dx.doi.org/10.3847/1538-4357/aaa40f},
year = {2018},
month = {jan},
publisher = {The American Astronomical Society},
volume = {853},
number = {2},
pages = {131},
author = {Lee, Bomee and Giavalisco, Mauro and Whitaker, Katherine and Williams, Christina C. and Ferguson, Henry C. and Acquaviva, Viviana and Koekemoer, Anton M. and Straughn, Amber N. and Guo, Yicheng and Kartaltepe, Jeyhan S. and Lotz, Jennifer and Pacifici, Camilla and Croton, Darren J. and Somerville, Rachel S. and Lu, Yu},
title = {The Intrinsic Characteristics of Galaxies on the SFR–M∗ Plane at 1.2 &lt; z &lt; 4: I. The Correlation between Stellar Age, Central Density, and Position Relative to the Main Sequence},
journal = {ApJ},
}

@article{Piotrowska2021,
    author = {Piotrowska, Joanna M and Bluck, Asa F L and Maiolino, Roberto and Peng, Yingjie},
    title = { On the quenching of star formation in observed and simulated central galaxies: evidence for the role of integrated AGN feedback},
    journal = {MNRAS},
    volume = {512},
    number = {1},
    pages = {1052-1090},
    year = {2021},
    month = {12},
    issn = {0035-8711},
    doi = {10.1093/mnras/stab3673},
    url = {https://doi.org/10.1093/mnras/stab3673},
    eprint = {https://academic.oup.com/mnras/article-pdf/512/1/1052/42960834/stab3673.pdf},
}

@article{Roberts69,
    author = {Roberts, M. S.},
    title = {Integral Properties of Spiral and Irregular Galaxies},
    journal = {AJ},
    year = {1969},
    volume = {74},
    number = {7},
    pages = {859-876},
    url = {doi.org/10.1086/110874},
    doi = {10.1086/110874}
}

@article{Baker21,
    author = {Baker, William M and Maiolino, Roberto and Bluck, Asa F L and Lin, Lihwai and Ellison, Sara L and Belfiore, Francesco and Pan, Hsi-An and Thorp, Mallory},
    title = {The ALMaQUEST survey IX: the nature of the resolved star forming main sequence},
    journal = {MNRAS},
    volume = {510},
    number = {3},
    pages = {3622-3628},
    year = {2021},
    month = {12},
    issn = {0035-8711},
    doi = {10.1093/mnras/stab3672},
    url = {https://doi.org/10.1093/mnras/stab3672},
    eprint = {https://academic.oup.com/mnras/article-pdf/510/3/3622/42151808/stab3672.pdf},
}

@article{Wang11,
    author = {Wang, Jing and Kauffmann, Guinevere and Overzier, Roderik and Catinella, Barbara and Schiminovich, David and Heckman, Timothy M. and Moran, Sean M. and Haynes, Martha P. and Giovanelli, Riccardo and Kong, Xu},
    title = {The GALEX Arecibo SDSS survey – III. Evidence for the inside-out formation of Galactic discs},
    journal = {MNRAS},
    volume = {412},
    number = {2},
    pages = {1081-1097},
    year = {2011},
    month = {03},
    issn = {0035-8711},
    doi = {10.1111/j.1365-2966.2010.17962.x},
    url = {https://doi.org/10.1111/j.1365-2966.2010.17962.x},
    eprint = {https://academic.oup.com/mnras/article-pdf/412/2/1081/5736593/mnras0412-1081.pdf},
}

@article{Saintonge11,
    author = {Saintonge, Amélie and Kauffmann, Guinevere and Kramer, Carsten and Tacconi, Linda J. and Buchbender, Christof and Catinella, Barbara and Fabello, Silvia and Graciá-Carpio, Javier and Wang, Jing and Cortese, Luca and Fu, Jian and Genzel, Reinhard and Giovanelli, Riccardo and Guo, Qi and Haynes, Martha P. and Heckman, Timothy M. and Krumholz, Mark R. and Lemonias, Jenna and Li, Cheng and Moran, Sean and Rodriguez-Fernandez, Nemesio and Schiminovich, David and Schuster, Karl and Sievers, Albrecht},
    title = {COLD GASS, an IRAM legacy survey of molecular gas in massive galaxies – I. Relations between H2, H i, stellar content and structural properties},
    journal = {MNRAS},
    volume = {415},
    number = {1},
    pages = {32-60},
    year = {2011},
    month = {07},
    issn = {0035-8711},
    doi = {10.1111/j.1365-2966.2011.18677.x},
    url = {https://doi.org/10.1111/j.1365-2966.2011.18677.x},
    eprint = {https://academic.oup.com/mnras/article-pdf/415/1/32/17328290/mnras0415-0032.pdf},
}

@article{Fisher13,
doi = {10.1088/0004-637X/764/2/174},
url = {https://dx.doi.org/10.1088/0004-637X/764/2/174},
year = {2013},
month = {feb},
publisher = {The American Astronomical Society},
volume = {764},
number = {2},
pages = {174},
author = {Fisher, David B. and Bolatto, Alberto and Drory, Niv and Combes, Francoise and Blitz, Leo and Wong, Tony},
title = {THE MOLECULAR GAS DENSITY IN GALAXY CENTERS AND HOW IT CONNECTS TO BULGES},
journal = {ApJ},
}

@article{Zhang2019,
author = {Zhang, Chengpeng and Peng, Yingjie and Ho, Luis C. and Maiolino, Roberto and Dekel, Avishai and Guo, Qi and Mannucci, Filippo and Li, Di and Yuan, Feng and Renzini, Alvio and Dou, Jing and Guo, Kexin and Man, Zhongyi and Li, Qiong},
doi = {10.3847/2041-8213/ab4ae4},
url = {https://dx.doi.org/10.3847/2041-8213/ab4ae4},
year = {2019},
month = {oct},
publisher = {The American Astronomical Society},
journal = {ApJL},
volume = {884},
number = {2},
pages = {L52},
}

@article{Leroy2008,
doi = {10.1088/0004-6256/136/6/2782},
url = {https://dx.doi.org/10.1088/0004-6256/136/6/2782},
year = {2008},
month = {nov},
publisher = {The American Astronomical Society},
volume = {136},
number = {6},
pages = {2782},
author = {Leroy, Adam K. and Walter, Fabian and Brinks, Elias and Bigiel, Frank and de Blok, W. J. G. and Madore, Barry and Thornley, M. D.},
title = {THE STAR FORMATION EFFICIENCY IN NEARBY GALAXIES: MEASURING WHERE GAS FORMS STARS EFFECTIVELY},
journal = {AJ},
}

@article{vanDokkum2015,
doi = {10.1088/0004-637X/813/1/23},
url = {https://dx.doi.org/10.1088/0004-637X/813/1/23},
year = {2015},
month = {oct},
publisher = {The American Astronomical Society},
volume = {813},
number = {1},
pages = {23},
author = {van Dokkum, Pieter G. and Nelson, Erica June and Franx, Marijn and Oesch, Pascal and Momcheva, Ivelina and Brammer, Gabriel and Schreiber, Natascha M. Förster and Skelton, Rosalind E. and Whitaker, Katherine E. and Wel, Arjen van der and Bezanson, Rachel and Fumagalli, Mattia and Illingworth, Garth D. and Kriek, Mariska and Leja, Joel and Wuyts, Stijn},
title = {FORMING COMPACT MASSIVE GALAXIES},
journal = {ApJ},
}

@article{Feigelson85,
    author = {Feigelson, E. D. and Nelson, P. I.},
    title = {Statistical methods for astronomical data with upper limits. I - Univariate distributions},
    journal = {ApJ},
    year = {1985},
    pages = {192-206},
    volume = {293}
}

@article{Fluetsch21,
    author = {Fluetsch, A and Maiolino, R and Carniani, S and Arribas, S and Belfiore, F and Bellocchi, E and Cazzoli, S and Cicone, C and Cresci, G and Fabian, A C and Gallagher, R and Ishibashi, W and Mannucci, F and Marconi, A and Perna, M and Sturm, E and Venturi, G},
    title = {Properties of the multiphase outflows in local (ultra)luminous infrared galaxies},
    journal = {MNRAS},
    volume = {505},
    number = {4},
    pages = {5753-5783},
    year = {2021},
    month = {06},
    issn = {0035-8711},
    doi = {10.1093/mnras/stab1666},
    url = {https://doi.org/10.1093/mnras/stab1666},
    eprint = {https://academic.oup.com/mnras/article-pdf/505/4/5753/42501756/stab1666.pdf},
}

@article{Garcia-Burillo2024,
	author = {García-Burillo, S. and Hicks, E. K. S. and Alonso-Herrero, A. and Pereira-Santaella, M. and Usero, A. and others},
	title = {Deciphering the imprint of active galactic nucleus feedback in Seyfert galaxies - Nuclear-scale molecular gas deficits},
	DOI= "10.1051/0004-6361/202450268",
	url= "https://doi.org/10.1051/0004-6361/202450268",
	journal = {A\&A},
	year = {2024},
	volume = {689},
	pages = "A347",
}

@article{Holden2024,
    author = {Holden, Luke R and Tadhunter, Clive and Audibert, Anelise and Oosterloo, Tom and Ramos Almeida, Cristina and Morganti, Raffaella and Pereira-Santaella, Miguel and Lamperti, Isabella},
    title = {ALMA reveals a compact and massive molecular outflow driven by the young AGN in a nearby ULIRG},
    journal = {MNRAS},
    volume = {530},
    number = {1},
    pages = {446-456},
    year = {2024},
    month = {03},
    issn = {0035-8711},
    doi = {10.1093/mnras/stae810},
    url = {https://doi.org/10.1093/mnras/stae810},
    eprint = {https://academic.oup.com/mnras/article-pdf/530/1/446/57192022/stae810.pdf},
}

@article{Dubois2013,
    author = {Dubois, Yohan and Gavazzi, Raphaël and Peirani, Sébastien and Silk, Joseph},
    title = {AGN-driven quenching of star formation: morphological and dynamical implications for early-type galaxies},
    journal = {MNRAS},
    volume = {433},
    number = {4},
    pages = {3297-3313},
    year = {2013},
    month = {06},
    issn = {0035-8711},
    doi = {10.1093/mnras/stt997},
    url = {https://doi.org/10.1093/mnras/stt997},
    eprint = {https://academic.oup.com/mnras/article-pdf/433/4/3297/4933412/stt997.pdf},
}

@article{Su2019,
   title={The failure of stellar feedback, magnetic fields, conduction, and morphological quenching in maintaining red galaxies},
   volume={487},
   ISSN={1365-2966},
   url={http://dx.doi.org/10.1093/mnras/stz1494},
   DOI={10.1093/mnras/stz1494},
   number={3},
   journal={MNRAS},
   publisher={Oxford University Press (OUP)},
   author={Su, Kung-Yi and Hopkins, Philip F and Hayward, Christopher C and Ma, Xiangcheng and Faucher-Giguère, Claude-André and Kereš, Dušan and Orr, Matthew E and Chan, T K and Robles, Victor H},
   year={2019},
   month=jun, 
   pages={4393–4408} 
}

@article{Ma2022,
doi = {10.3847/1538-4357/aca326},
url = {https://dx.doi.org/10.3847/1538-4357/aca326},
year = {2022},
month = {dec},
publisher = {The American Astronomical Society},
volume = {941},
number = {2},
pages = {205},
author = {Ma, Wenlin and Liu, Kexin and Guo, Hong and Cui, Weiguang and Jones, Michael G. and Wang, Jing and Zhang, Le and Davé, Romeel},
title = {Effects of Active Galactic Nucleus Feedback on Cold Gas Depletion and Quenching of Central Galaxies},
journal = {ApJ},
}

@article{Guo2021,
doi = {10.3847/1538-4357/abf115},
url = {https://dx.doi.org/10.3847/1538-4357/abf115},
year = {2021},
month = {jun},
publisher = {The American Astronomical Society},
volume = {914},
number = {1},
pages = {7},
author = {Guo, Yicheng and Carleton, Timothy and Bell, Eric F. and Chen, Zhu and Dekel, Avishai and Faber, S. M. and Giavalisco, Mauro and Kocevski, Dale D. and Koekemoer, Anton M. and Koo, David C. and Kurczynski, Peter and Lee, Seong-Kook and Liu, F. S. and Papovich, Casey and Pérez-González, Pablo G.},
title = {Implications of Increased Central Mass Surface Densities for the Quenching of Low-mass Galaxies},
journal = {ApJ}
}

@article{Croton_2006,
   title={The many lives of active galactic nuclei: cooling flows, black holes and the luminosities and colours of galaxies},
   volume={365},
   ISSN={1365-2966},
   url={http://dx.doi.org/10.1111/j.1365-2966.2005.09675.x},
   DOI={10.1111/j.1365-2966.2005.09675.x},
   number={1},
   journal={MNRAS},
   publisher={Oxford University Press (OUP)},
   author={Croton, Darren J. and Springel, Volker and White, Simon D. M. and De Lucia, G. and Frenk, C. S. and Gao, L. and Jenkins, A. and Kauffmann, G. and Navarro, J. F. and Yoshida, N.},
   year={2006},
   month=jan, pages={11–28} }

@article{Baldwin81,
    author = {Baldwin, J. A. and Phillips, M. M. and Terlevich, R.},
    title = {Classification parameters for the emission-line spectra of extragalactic objects.},
    journal = {PASP},
    year = {1981},
    volume = {93},
    pages = {5-19},
    url = {https://iopscience.iop.org/article/10.1086/130766},
    doi = {10.1086/130766}
}

@article{Gereb_2015,
    author = {K. Ger\'eb and R. Morganti and T. A. Oosterloo and L. Hoppmann and L. Staveley-Smith},
    title = {From star-forming galaxies to AGN: the global HI content from a stacking experiment},
    journal = {A\&A},
    year = {2015},
    volume = {580},
    pages = {A43},
    url = {	https://doi.org/10.1051/0004-6361/201424810},
    doi = {10.1051/0004-6361/201424810}
}

@article{Fabello_2011,
    author = {Fabello, Silvia and Kauffmann, Guinevere and Catinella, Barbara and Giovanelli, Riccardo and Haynes, Martha P. and Heckman, Timothy M. and Schiminovich, David},
    title = {Arecibo Legacy Fast ALFA H i data stacking – II. H i content of the host galaxies of active galactic nuclei},
    journal = {MNRAS},
    volume = {416},
    number = {3},
    pages = {1739-1744},
    year = {2011},
    month = {09},
    issn = {0035-8711},
    doi = {10.1111/j.1365-2966.2011.18825.x},
    url = {https://doi.org/10.1111/j.1365-2966.2011.18825.x},
    eprint = {https://academic.oup.com/mnras/article-pdf/416/3/1739/2840189/mnras0416-1739.pdf},
}

@article{SciPy20,
  author  = {Virtanen, Pauli T and Gommers, Ralf and Oliphant, Travis E and Haberland, Matt and Reddy, Tyler and Cournapeau, David and Burovski, Evgeni and Peterson, Warren and Weckesser, Warren and Bright, John and {van der Walt}, St{\'e}fan J and Brett, Matthew and Wilson, Joshua and Millman, K Jarrod and Mayorov, Nikolay and Nelson, Andrew RJ and Jones, Eric and Bezanson, Jeff and van den Berg, Pearu and Ivanov, Sergey and Brumback, Andrzej and Fitzgerald, Christopher J and Gilbert, Nathaniel and Gomersall, Thomas and Redmond, David SA and O'Leary, Paul and {Hagen}, Erik and Mason, Jonathan D and Schep, Sebastian and Villa, Crist{\'o}bal and Turrell, Bart{\'o}miej and Manohar, Shivam and Carr, Kyle and SciPy 1. 0 Contributors},
  title   = {SciPy 1.0: fundamental algorithms for scientific computing in Python},
  journal = {Nature Methods},
  volume  = {17},
  number  = {3},
  pages = {261--272},
  year    = {2020},
  doi = {10.1038/s41592-019-0686-2}
}

@article{astropy13,
Adsurl = {http://adsabs.harvard.edu/abs/2013A%26A...558A..33A},
Archiveprefix = {arXiv},
Author = {{Astropy Collaboration} and {Robitaille}, T.~P. and {Tollerud}, E.~J. and {Greenfield}, P. and {Droettboom}, M. and {Bray}, E. and {Aldcroft}, T. and {Davis}, M. and {Ginsburg}, A. and {Price-Whelan}, A.~M. and {Kerzendorf}, W.~E. and {Conley}, A. and {Crighton}, N. and {Barbary}, K. and {Muna}, D. and {Ferguson}, H. and {Grollier}, F. and {Parikh}, M.~M. and {Nair}, P.~H. and {Unther}, H.~M. and {Deil}, C. and {Woillez}, J. and {Conseil}, S. and {Kramer}, R. and {Turner}, J.~E.~H. and {Singer}, L. and {Fox}, R. and {Weaver}, B.~A. and {Zabalza}, V. and {Edwards}, Z.~I. and {Azalee Bostroem}, K. and {Burke}, D.~J. and {Casey}, A.~R. and {Crawford}, S.~M. and {Dencheva}, N. and {Ely}, J. and {Jenness}, T. and {Labrie}, K. and {Lim}, P.~L. and {Pierfederici}, F. and {Pontzen}, A. and {Ptak}, A. and {Refsdal}, B. and {Servillat}, M. and {Streicher}, O.},
Doi = {10.1051/0004-6361/201322068},
Eid = {A33},
Eprint = {1307.6212},
Journal = {\aap},
Month = oct,
Pages = {A33},
Primaryclass = {astro-ph.IM},
Title = {{Astropy: A community Python package for astronomy}},
Volume = 558,
Year = 2013}

@ARTICLE{astropy18,
       author = {{Astropy Collaboration} and {Price-Whelan}, A.~M. and
         {Sip{\H{o}}cz}, B.~M. and {G{\"u}nther}, H.~M. and {Lim}, P.~L. and
         {Crawford}, S.~M. and {Conseil}, S. and {Shupe}, D.~L. and
         {Craig}, M.~W. and {Dencheva}, N. and {Ginsburg}, A. and {Vand
        erPlas}, J.~T. and {Bradley}, L.~D. and {P{\'e}rez-Su{\'a}rez}, D. and
         {de Val-Borro}, M. and {Aldcroft}, T.~L. and {Cruz}, K.~L. and
         {Robitaille}, T.~P. and {Tollerud}, E.~J. and {Ardelean}, C. and
         {Babej}, T. and {Bach}, Y.~P. and {Bachetti}, M. and {Bakanov}, A.~V. and
         {Bamford}, S.~P. and {Barentsen}, G. and {Barmby}, P. and
         {Baumbach}, A. and {Berry}, K.~L. and {Biscani}, F. and {Boquien}, M. and
         {Bostroem}, K.~A. and {Bouma}, L.~G. and {Brammer}, G.~B. and
         {Bray}, E.~M. and {Breytenbach}, H. and {Buddelmeijer}, H. and
         {Burke}, D.~J. and {Calderone}, G. and {Cano Rodr{\'\i}guez}, J.~L. and
         {Cara}, M. and {Cardoso}, J.~V.~M. and {Cheedella}, S. and {Copin}, Y. and
         {Corrales}, L. and {Crichton}, D. and {D'Avella}, D. and {Deil}, C. and
         {Depagne}, {\'E}. and {Dietrich}, J.~P. and {Donath}, A. and
         {Droettboom}, M. and {Earl}, N. and {Erben}, T. and {Fabbro}, S. and
         {Ferreira}, L.~A. and {Finethy}, T. and {Fox}, R.~T. and
         {Garrison}, L.~H. and {Gibbons}, S.~L.~J. and {Goldstein}, D.~A. and
         {Gommers}, R. and {Greco}, J.~P. and {Greenfield}, P. and
         {Groener}, A.~M. and {Grollier}, F. and {Hagen}, A. and {Hirst}, P. and
         {Homeier}, D. and {Horton}, A.~J. and {Hosseinzadeh}, G. and {Hu}, L. and
         {Hunkeler}, J.~S. and {Ivezi{\'c}}, {\v{Z}}. and {Jain}, A. and
         {Jenness}, T. and {Kanarek}, G. and {Kendrew}, S. and {Kern}, N.~S. and
         {Kerzendorf}, W.~E. and {Khvalko}, A. and {King}, J. and {Kirkby}, D. and
         {Kulkarni}, A.~M. and {Kumar}, A. and {Lee}, A. and {Lenz}, D. and
         {Littlefair}, S.~P. and {Ma}, Z. and {Macleod}, D.~M. and
         {Mastropietro}, M. and {McCully}, C. and {Montagnac}, S. and
         {Morris}, B.~M. and {Mueller}, M. and {Mumford}, S.~J. and {Muna}, D. and
         {Murphy}, N.~A. and {Nelson}, S. and {Nguyen}, G.~H. and
         {Ninan}, J.~P. and {N{\"o}the}, M. and {Ogaz}, S. and {Oh}, S. and
         {Parejko}, J.~K. and {Parley}, N. and {Pascual}, S. and {Patil}, R. and
         {Patil}, A.~A. and {Plunkett}, A.~L. and {Prochaska}, J.~X. and
         {Rastogi}, T. and {Reddy Janga}, V. and {Sabater}, J. and
         {Sakurikar}, P. and {Seifert}, M. and {Sherbert}, L.~E. and
         {Sherwood-Taylor}, H. and {Shih}, A.~Y. and {Sick}, J. and
         {Silbiger}, M.~T. and {Singanamalla}, S. and {Singer}, L.~P. and
         {Sladen}, P.~H. and {Sooley}, K.~A. and {Sornarajah}, S. and
         {Streicher}, O. and {Teuben}, P. and {Thomas}, S.~W. and
         {Tremblay}, G.~R. and {Turner}, J.~E.~H. and {Terr{\'o}n}, V. and
         {van Kerkwijk}, M.~H. and {de la Vega}, A. and {Watkins}, L.~L. and
         {Weaver}, B.~A. and {Whitmore}, J.~B. and {Woillez}, J. and
         {Zabalza}, V. and {Astropy Contributors}},
        title = "{The Astropy Project: Building an Open-science Project and Status of the v2.0 Core Package}",
      journal = {\aj},
         year = 2018,
        month = sep,
       volume = {156},
       number = {3},
          eid = {123},
        pages = {123},
          doi = {10.3847/1538-3881/aabc4f},
archivePrefix = {arXiv},
       eprint = {1801.02634},
 primaryClass = {astro-ph.IM},
       adsurl = {https://ui.adsabs.harvard.edu/abs/2018AJ....156..123A}
}

@ARTICLE{astropy22,
       author = {{Astropy Collaboration} and {Price-Whelan}, Adrian M. and {Lim}, Pey Lian and {Earl}, Nicholas and {Starkman}, Nathaniel and {Bradley}, Larry and {Shupe}, David L. and {Patil}, Aarya A. and {Corrales}, Lia and {Brasseur}, C.~E. and {N{"o}the}, Maximilian and {Donath}, Axel and {Tollerud}, Erik and {Morris}, Brett M. and {Ginsburg}, Adam and {Vaher}, Eero and {Weaver}, Benjamin A. and {Tocknell}, James and {Jamieson}, William and {van Kerkwijk}, Marten H. and {Robitaille}, Thomas P. and {Merry}, Bruce and {Bachetti}, Matteo and {G{"u}nther}, H. Moritz and {Aldcroft}, Thomas L. and {Alvarado-Montes}, Jaime A. and {Archibald}, Anne M. and {B{'o}di}, Attila and {Bapat}, Shreyas and {Barentsen}, Geert and {Baz{'a}n}, Juanjo and {Biswas}, Manish and {Boquien}, M{'e}d{'e}ric and {Burke}, D.~J. and {Cara}, Daria and {Cara}, Mihai and {Conroy}, Kyle E. and {Conseil}, Simon and {Craig}, Matthew W. and {Cross}, Robert M. and {Cruz}, Kelle L. and {D'Eugenio}, Francesco and {Dencheva}, Nadia and {Devillepoix}, Hadrien A.~R. and {Dietrich}, J{"o}rg P. and {Eigenbrot}, Arthur Davis and {Erben}, Thomas and {Ferreira}, Leonardo and {Foreman-Mackey}, Daniel and {Fox}, Ryan and {Freij}, Nabil and {Garg}, Suyog and {Geda}, Robel and {Glattly}, Lauren and {Gondhalekar}, Yash and {Gordon}, Karl D. and {Grant}, David and {Greenfield}, Perry and {Groener}, Austen M. and {Guest}, Steve and {Gurovich}, Sebastian and {Handberg}, Rasmus and {Hart}, Akeem and {Hatfield-Dodds}, Zac and {Homeier}, Derek and {Hosseinzadeh}, Griffin and {Jenness}, Tim and {Jones}, Craig K. and {Joseph}, Prajwel and {Kalmbach}, J. Bryce and {Karamehmetoglu}, Emir and {Ka{l}uszy{'n}ski}, Miko{l}aj and {Kelley}, Michael S.~P. and {Kern}, Nicholas and {Kerzendorf}, Wolfgang E. and {Koch}, Eric W. and {Kulumani}, Shankar and {Lee}, Antony and {Ly}, Chun and {Ma}, Zhiyuan and {MacBride}, Conor and {Maljaars}, Jakob M. and {Muna}, Demitri and {Murphy}, N.~A. and {Norman}, Henrik and {O'Steen}, Richard and {Oman}, Kyle A. and {Pacifici}, Camilla and {Pascual}, Sergio and {Pascual-Granado}, J. and {Patil}, Rohit R. and {Perren}, Gabriel I. and {Pickering}, Timothy E. and {Rastogi}, Tanuj and {Roulston}, Benjamin R. and {Ryan}, Daniel F. and {Rykoff}, Eli S. and {Sabater}, Jose and {Sakurikar}, Parikshit and {Salgado}, Jes{'u}s and {Sanghi}, Aniket and {Saunders}, Nicholas and {Savchenko}, Volodymyr and {Schwardt}, Ludwig and {Seifert-Eckert}, Michael and {Shih}, Albert Y. and {Jain}, Anany Shrey and {Shukla}, Gyanendra and {Sick}, Jonathan and {Simpson}, Chris and {Singanamalla}, Sudheesh and {Singer}, Leo P. and {Singhal}, Jaladh and {Sinha}, Manodeep and {Sip{H{o}}cz}, Brigitta M. and {Spitler}, Lee R. and {Stansby}, David and {Streicher}, Ole and {Sumak}, Jani and {Swinbank}, John D. and {Taranu}, Dan S. and {Tewary}, Nikita and {Tremblay}, Grant R. and {Val-Borro}, Miguel de and {Van Kooten}, Samuel J. and {Vasovi{'c}}, Zlatan and {Verma}, Shresth and {de Miranda Cardoso}, Jos{'e} Vin{'i}cius and {Williams}, Peter K.~G. and {Wilson}, Tom J. and {Winkel}, Benjamin and {Wood-Vasey}, W.~M. and {Xue}, Rui and {Yoachim}, Peter and {Zhang}, Chen and {Zonca}, Andrea and {Astropy Project Contributors}},
        title = "{The Astropy Project: Sustaining and Growing a Community-oriented Open-source Project and the Latest Major Release (v5.0) of the Core Package}",
      journal = {\apj},
         year = 2022,
        month = aug,
       volume = {935},
       number = {2},
          eid = {167},
        pages = {167},
          doi = {10.3847/1538-4357/ac7c74},
archivePrefix = {arXiv},
       eprint = {2206.14220},
 primaryClass = {astro-ph.IM},
       adsurl = {https://ui.adsabs.harvard.edu/abs/2022ApJ...935..167A}
}

@software{pandas20,
    author       = {The pandas development team},
    title        = {pandas-dev/pandas: Pandas},
    month        = feb,
    year         = 2020,
    publisher    = {Zenodo},
    version      = {latest},
    doi          = {10.5281/zenodo.3509134},
    url          = {https://doi.org/10.5281/zenodo.3509134}
}

@Article{Hunter07,
  Author    = {Hunter, J. D.},
  Title     = {Matplotlib: A 2D graphics environment},
  Journal   = {Computing in Science \& Engineering},
  Volume    = {9},
  Number    = {3},
  Pages     = {90--95},
  publisher = {IEEE COMPUTER SOC},
  doi       = {10.1109/MCSE.2007.55},
  year      = 2007
}

@article{Waskom21,
    doi = {10.21105/joss.03021},
    url = {https://doi.org/10.21105/joss.03021},
    year = {2021},
    publisher = {The Open Journal},
    volume = {6},
    number = {60},
    pages = {3021},
    author = {Michael L. Waskom},
    title = {seaborn: statistical data visualization},
    journal = {Journal of Open Source Software}
 }

@Article{         Harris20,
 title         = {Array programming with {NumPy}},
 author        = {Charles R. Harris and K. Jarrod Millman and St{\'{e}}fan J.
                 van der Walt and Ralf Gommers and Pauli Virtanen and David
                 Cournapeau and Eric Wieser and Julian Taylor and Sebastian
                 Berg and Nathaniel J. Smith and Robert Kern and Matti Picus
                 and Stephan Hoyer and Marten H. van Kerkwijk and Matthew
                 Brett and Allan Haldane and Jaime Fern{\'{a}}ndez del
                 R{\'{i}}o and Mark Wiebe and Pearu Peterson and Pierre
                 G{\'{e}}rard-Marchant and Kevin Sheppard and Tyler Reddy and
                 Warren Weckesser and Hameer Abbasi and Christoph Gohlke and
                 Travis E. Oliphant},
 year          = {2020},
 month         = sep,
 journal       = {Nature},
 volume        = {585},
 number        = {7825},
 pages         = {357--362},
 doi           = {10.1038/s41586-020-2649-2},
 publisher     = {Springer Science and Business Media {LLC}},
 url           = {https://doi.org/10.1038/s41586-020-2649-2}
}

@article{Moles_1995,
    author = {Moles, Mariano and M\'arquez, Isabel and P\'erez, Enrique},
    title = {The relation between dynamical perturbations, morphology, and nuclear activity in spiral galaxies},
    journal = {ApJ},
    year = {1995},
    volume = {438},
    pages = {604-609},
}

@article{Salvestrini_2022,
	author = {Salvestrini, F. and Gruppioni, C. and Hatziminaoglou, E. and Pozzi, F. and Vignali, C. and Casasola, V. and Paladino, R. and Aalto, S. and Andreani, P. and Marchesi, S. and Stanke, T.},
	title = {The molecular gas properties in local Seyfert 2 galaxies⋆},
	DOI= "10.1051/0004-6361/202142760",
	url= "https://doi.org/10.1051/0004-6361/202142760",
	journal = {A\&A},
	year = 2022,
	volume = 663,
	pages = "A28",
}

@article{Hunt_1999,
doi = {10.1086/306607},
url = {https://dx.doi.org/10.1086/306607},
year = {1999},
month = {jan},
publisher = {},
volume = {510},
number = {2},
pages = {637},
author = {Hunt, L. K. and Malkan, M. A. and Moriondo, G. and Salvati, M.},
title = {The Disks of Galaxies with Seyfert and Starburst Nuclei. II. Near-Infrared Structural Properties},
journal = {ApJ},
}

@article{Singh_2013,
	author = {Singh, R. and van de Ven, G. and Jahnke, K. and Lyubenova, M. and Falcón-Barroso, J. and Alves, J. and Cid Fernandes, R. and {Galbany, L.} and {García-Benito, R.} and {Husemann, B.} and {Kennicutt, R. C.} and {Marino, R. A.} and {Márquez, I.} and {Masegosa, J.} and {Mast, D.} and {Pasquali, A.} and {Sánchez, S. F.} and {Walcher, J.} and {Wild, V.} and {Wisotzki, L.} and {Ziegler, B.} and {the CALIFA collaboration}},
	title = {The nature of LINER galaxies: - Ubiquitous hot old stars and rare accreting black holes},
	DOI= "10.1051/0004-6361/201322062",
	url= "https://doi.org/10.1051/0004-6361/201322062",
	journal = {A\&A},
	year = 2013,
	volume = 558,
	pages = "A43",
	month = "",
}

@ARTICLE{Marquez_2017,
AUTHOR={M\'arquez, Isabel  and Masegosa, Josefa  and Gonz\'alez-Martin, Omaira  and Hern`'andez-Garcia, Lorena  and Povi\`c, Mirjana  and Netzer, Hagai  and Cazzoli, Sara  and del Olmo, Ascensión },      
TITLE={The AGN Nature of LINER Nuclear Sources},      
JOURNAL={FSPAS},  
VOLUME={4},
YEAR={2017},
URL={https://www.frontiersin.org/journals/astronomy-and-space-sciences/articles/10.3389/fspas.2017.00034},
DOI={10.3389/fspas.2017.00034},
ISSN={2296-987X},
}

@software{davidson_pilon_2024_14007206,
  author       = {Davidson-Pilon, Cameron},
  title        = {lifelines, survival analysis in Python},
  month        = oct,
  year         = 2024,
  publisher    = {Zenodo},
  version      = {v0.30.0},
  doi          = {10.5281/zenodo.14007206},
  url          = {https://doi.org/10.5281/zenodo.14007206},
}

@article{Gatto2024,
    author = {Gatto, Lara and Storchi-Bergmann, T and Riffel, Rogemar A and Riffel, Rogério and Rembold, Sandro B and Schimoia, Jaderson S and Mallmann, Nicolas D and Ilha, Gabriele S},
    title = {The extent and power of ‘maintenance mode’ feedback in MaNGA AGN},
    journal = {MNRAS},
    volume = {530},
    number = {3},
    pages = {3059-3074},
    year = {2024},
    month = {04},
    issn = {0035-8711},
    doi = {10.1093/mnras/stae989},
    url = {https://doi.org/10.1093/mnras/stae989},
    eprint = {https://academic.oup.com/mnras/article-pdf/530/3/3059/57362616/stae989.pdf},
}

@article{WHAN_2011,
    author = {Cid Fernandes, R. and Stasińska, G. and Mateus, A. and Vale Asari, N.},
    title = {A comprehensive classification of galaxies in the Sloan Digital Sky Survey: how to tell true from fake AGN?},
    journal = {MNRAS},
    volume = {413},
    number = {3},
    pages = {1687-1699},
    year = {2011},
    month = {05},
    issn = {0035-8711},
    doi = {10.1111/j.1365-2966.2011.18244.x},
    url = {https://doi.org/10.1111/j.1365-2966.2011.18244.x},
    eprint = {https://academic.oup.com/mnras/article-pdf/413/3/1687/2873756/mnras0413-1687.pdf},
}

@article{WHAN_2010,
    author = {Fernandes, R. Cid and Stasińska, G. and Schlickmann, M. S. and Mateus, A. and Asari, N. Vale and Schoenell, W. and Sodré, L., Jr and (the SEAGal collaboration)},
    title = {Alternative diagnostic diagrams and the ‘forgotten’ population of weak line galaxies in the SDSS},
    journal = {MNRAS},
    volume = {403},
    number = {2},
    pages = {1036-1053},
    year = {2010},
    month = {03},
    issn = {0035-8711},
    doi = {10.1111/j.1365-2966.2009.16185.x},
    url = {https://doi.org/10.1111/j.1365-2966.2009.16185.x},
    eprint = {https://academic.oup.com/mnras/article-pdf/403/2/1036/4012938/mnras0403-1036.pdf},
}

@article{Khan_2025,
doi = {10.3847/1538-4357/ade99e},
url = {https://dx.doi.org/10.3847/1538-4357/ade99e},
year = {2025},
month = {aug},
publisher = {The American Astronomical Society},
volume = {990},
number = {1},
pages = {6},
author = {Khan, Fazeel Mahmood and Rodríguez, Ángel and Macció, Andrea V. and Sharma, Smarika and Cho, Changhyun},
title = {Active Galactic Nucleus Feedback-induced Stellar Density Expansion in the Inner Regions of Early-type Galaxies},
journal = {ApJ},
}

@article{LINMIX2007,
doi = {10.1086/519947},
url = {https://doi.org/10.1086/519947},
year = {2007},
month = {aug},
publisher = {},
volume = {665},
number = {2},
pages = {1489},
author = {Kelly, Brandon C.},
title = {Some Aspects of Measurement Error in Linear Regression of Astronomical Data},
journal = {ApJ},
}

@article{McConnell_Ma_2013,
doi = {10.1088/0004-637X/764/2/184},
url = {https://doi.org/10.1088/0004-637X/764/2/184},
year = {2013},
month = {feb},
publisher = {The American Astronomical Society},
volume = {764},
number = {2},
pages = {184},
author = {McConnell, Nicholas J. and Ma, Chung-Pei},
title = {REVISITING THE SCALING RELATIONS OF BLACK HOLE MASSES AND HOST GALAXY PROPERTIES},
journal = {ApJ},
}

@article{Bentz_2013,
doi = {10.1088/0004-637X/767/2/149},
url = {https://doi.org/10.1088/0004-637X/767/2/149},
year = {2013},
month = {apr},
publisher = {The American Astronomical Society},
volume = {767},
number = {2},
pages = {149},
author = {Bentz, Misty C. and Denney, Kelly D. and Grier, Catherine J. and Barth, Aaron J. and Peterson, Bradley M. and Vestergaard, Marianne and Bennert, Vardha N. and Canalizo, Gabriela and De Rosa, Gisella and Filippenko, Alexei V. and Gates, Elinor L. and Greene, Jenny E. and Li, Weidong and Malkan, Matthew A. and Pogge, Richard W. and Stern, Daniel and Treu, Tommaso and Woo, Jong-Hak},
title = {THE LOW-LUMINOSITY END OF THE RADIUS–LUMINOSITY RELATIONSHIP FOR ACTIVE GALACTIC NUCLEI},
journal = {ApJ},
}

@article{Hagedorn2024,
	author = {Hagedorn, B. and Cicone, C. and Sarzi, M. and Saintonge, A. and Severgnini, P. and Vignali, C. and Shen, S. and Rubinur, K. and Schimek, A. and Lasrado, A.},
	title = {Molecular gas scaling relations for local star-forming galaxies in the low-M* regime},
	DOI= "10.1051/0004-6361/202449773",
	url= "https://doi.org/10.1051/0004-6361/202449773",
	journal = {A\&A},
	year = 2024,
	volume = 687,
	pages = "A244",
}

@article{Hickox2018,
doi = {10.1146/annurev-astro-081817-051803},
url = {https://www.annualreviews.org/content/journals/10.1146/annurev-astro-081817-051803},
year = {2018},
month = {sep},
publisher = {Annual Review of Astronomy and Astrophysics},
volume = {56},
pages = {625},
author = {Hickox, Ryan C. and Alexander, David M.},
title = {Obscured Active Galactic Nuclei},
journal = {ARAA},
}

@article{Hopkins2008,
doi = {10.1086/524362},
url = {https://iopscience.iop.org/article/10.1086/524362},
year = {2008},
month = {april},
publisher = {The American Astronomical Society},
volume = {175},
number = {2},
pages = {356},
author = {Hopkins, Philip F. and Hernquist, Lars and Cox, Thomas J. and Kereš, Dušan},
title = {A Cosmological Framework for the Co-Evolution of Quasars, Supermassive Black Holes, and Elliptical Galaxies. I. Galaxy Mergers and Quasar Activity},
journal = {ApJS},
}

@article{Praestgaard1995,
  title={Permutation and bootstrap Kolmogorov-Smirnov tests for the equality of two distributions},
  author={Praestgaard, Jens Thomas},
  journal={Scandinavian Journal of Statistics},
  pages={305-322},
  year={1995},
  publisher={JSTOR}
}
\bibliographystyle{aasjournal}

\end{document}